\documentclass[fleqn,usenatbib]{mnras}

\usepackage[T1]{fontenc}
\usepackage{ae,aecompl}

\usepackage{graphicx}	
\usepackage{amsmath}	
\usepackage{amssymb}	

\newcommand{\Ms}{{\ensuremath{\mathrm{M}_{\sun}}}}
\newcommand{\Ni}{{\ensuremath{\dot{N}_{\mathrm{ion}}}}}

\title[Pop~III Survivors
]{Predicting the locations of possible long-lived low-mass first stars: Importance of satellite dwarf galaxies}

\author[M. Magg et al.]{Mattis Magg$^{1}$\thanks{E-mail: mattis.magg@stud.uni-heidelberg.de}, 
Tilman Hartwig$^{2, 3}$, Bhaskar Agarwal$^{1}$, Anna Frebel$^{4}$,\newauthor Simon C. O. Glover$^{1}$, Brendan F. Griffen$^{4}$ and Ralf S. Klessen$^{1,5}$\\
$^{1}$Universit\"at Heidelberg, Zentrum f\"ur Astronomie, Institut f\"ur Theoretische Astrophysik, Heidelberg, Germany\\
$^{2}$Sorbonne Universit\'es, UPMC Univ Paris 06, UMR 7095, Institut d'Astrophysique de Paris, Paris, France\\
$^{3}$CNRS, UMR 7095, Institut d'Astrophysique de Paris, Paris, France\\
$^{4}$Department of Physics \& Kavli Institute for Astrophysics and Space Research, Massachusetts Institute of Technology, Cambridge, USA\\
$^{5}$Universit\"{a}t Heidelberg, Interdisziplin\"{a}res Zentrum f\"{u}r Wissenschaftliches Rechnen, Heidelberg, Germany\\
}

\date{Accepted XXX. Received YYY; in original form ZZZ}

\pubyear{2017}

\begin{document}
\label{firstpage}
\pagerange{\pageref{firstpage}--\pageref{lastpage}}
\maketitle

\begin{abstract}
The search for metal-free stars has so far been unsuccessful, proving that if there are surviving stars from the first generation, they are rare, they have been polluted, or we have been looking in the wrong place. To predict the likely location of Population~III (Pop~III) survivors, we semi-analytically model early star formation in progenitors of Milky Way-like galaxies and their environments. We base our model on merger trees from the high-resolution dark matter only simulation suite \textit{Caterpillar}. Radiative and chemical feedback are taken into account self-consistently, based on the spatial distribution of the haloes. Our results are consistent with the non-detection of Pop~III survivors in the Milky Way today. We find that possible surviving Population III stars are more common in Milky Way satellites than in the main Galaxy. In particular, low mass Milky Way satellites contain a much larger fraction of Pop~III stars than the Milky Way. Such nearby, low mass Milky Way satellites are promising targets for future attempts to find Pop~III survivors, especially for high-resolution, high signal-to-noise spectroscopic observations. We provide the probabilities of finding a Pop~III survivor in the red giant branch phase for all known Milky Way satellites to guide future observations.
\end{abstract}




\section{INTRODUCTION}
The appearance of the first stars marked a primary transition in cosmic history. Their light ended the so-called `dark ages' and they played a key role in cosmic metal enrichment and reionization. This regulated early galaxy formation, and thus shaped the galaxies we still see today. Studying stellar birth in the primordial Universe and the first galaxies is a relatively young discipline of astrophysical research. It heavily relies on theoretical model building and numerical simulations. To a large degree, this is due to the fact that, unlike in present-day star formation, stringent observational constraints are rare and extremely difficult to obtain. 

The first generation of stars, the so-called Population~III (or Pop~III) formed from truly metal-free primordial gas. They have long been thought to emerge in isolation with only one massive star of about $100\,${\Ms} in the centre of a dark matter halo \cite[e.g.][]{omukai2001,abn02, bcl02, tm04,yoha06, Oshea07}. In the past ten years or so, however, this picture has seen substantial revision. Most current models find that the protostellar accretion disks around the first stars are highly susceptible to fragmentation, and as a consequence Pop~III stars typically form as members of multiple stellar systems with a wide distribution of masses, possibly ranging from the sub-stellar regime up to several $100\,{\Ms}$ \cite[e.g.][]{machida2008, turk09,Clark11,Greif11b, get12, Dopcke13,Stacy14,Stacy10,Stacy16}. Further complexity is revealed in studies that include radiative feedback \citep{sm11,hos11,Hirano14,hir15,Hosokawa16}, magnetic fields \citep{machida2006, Peters2014}, dark matter annihilation \citep{ripa10, Smith2012, Stacy2014}, or the influence of primordial streaming velocities \citep{th10,Stacy11,Greif11a, Maio11, naoz13}.

As mass is the primary parameter that determines the stellar lifetime, energy production, and nucleosynthetic yields, identifying and characterizing the physical processes that determine the initial mass function (IMF) is a prerequisite for our understanding of Pop~III star formation. To make progress, our current theoretical models need clear guidance from the observations.

Unfortunately, it is impossible to directly observe Pop~III stars in the high-redshift Universe in order to collect any observational constraints, as they are much too faint to be within reach of even the next generation of space telescopes, such as the James Webb Space Telescope. Supernova explosions occurring at the end of the lives of massive Pop~III stars should be detectable \citep[see e.g.][]{kasen11,wet13c,wet13b} but provide few constraints on the Pop~III IMF \citep{Magg16}, and none whatsoever on the form of the IMF below $8\,\Ms$.

Current Galactic archaeology studies \citep{bc05,cau13,Frebel15} in the halo and bulge of our Milky Way or the analysis of stars in nearby satellite dwarf galaxies \citep{Kirby15,Ji16,Skuladottir17} can contribute to our knowledge of primordial stars in two ways. First, the chemical abundance patterns measured in ancient, extremely metal-poor stars can be used to infer the properties of their progenitor stars which provided the heavy elements incorporated into the observed stars \cite[][]{hw02, hw10}. Assuming that the gas from which the oldest and most metal-poor stars in the Galaxy formed has been enriched by one or at most two supernovae explosions, the measured relative abundances of heavy elements in these stars are most consistent with progenitor core collapse supernovae from Pop~III stars in the mass range of several tens of solar masses, but below 100 \Ms\ \cite[e.g.][]{ fet05, Iwamoto2005, Lai2008, jet09a, jet10, caffau12, nor13,keller14,cm14,ishigaki14,Placco14,bf15, Bessell15}. Together with the fact that no genuine signatures of pair-instability supernovae from massive stars in the range of $\sim 140$ to $260\,${\Ms} have been found, this places constraints on the intermediate and high-mass end of the primordial IMF \citep[but see][]{aoki14}. Second, the theory of stellar evolution \cite[e.g.][]{kippenhahn2012} tells us that any low-mass stars with $0.8\,${\Ms} or less can survive until the present day if they formed early in the universe. This would also be the case for any low-mass Pop~III stars. Hence, if these stars ever existed, then they should eventually be detected in detailed Local Group surveys. 

So far, no Pop~III candidate has been found. The closest to it is a metal-poor star with an iron abundance\footnote{For the relative abundance of two elements A and B we use the notation $\mbox{[A/B]} = \log_{10}(m_\mathrm{A}/m_\mathrm{A})-\log_{10} (m_{\mathrm{A},\sun} /m_{\mathrm{b},\sun})$ where $m_{\mathrm{A}}$ and $m_{\mathrm{B}}$ are the abundances of element A and B, and $m_{\mathrm{A},\sun}$ and $m_{\mathrm{A},\sun}$ are their solar abundances.}
of $\mbox{[Fe/H]}<-7$\ \citep{keller14} and a carbon abundance orders of magnitude higher than that of iron. This, together with an abundance of $\mbox{[Ca/H]}\sim-7$, suggests that these elements formed in a high-mass Pop~III star. Another close example is a star with a total metal abundance of $\mbox{[M/H]}\sim-5$ \citep{caffau12}. As existing surveys are further exploited for more of the most metal-poor stars (e.g. ToPOS, \citealt{cau13}; SkyMapper; \citealt{Jacobson15}), any potential Pop~III star should be detectable as well. Search methods are usually based on the strength of the Ca\,II\,K line. An absence of this line would be a first indicator of a Pop~III star and is well within the technical reach of current surveys. An alternative is to search for Pop~III stars in the Milky Way's dwarf galaxies. However, since stars are faint and only the brightest can be observed, the number of investigated stars remains relatively small, especially in the dimmest ultra-faint dwarfs. While all stars in these systems are metal-poor and have $\mbox{[Fe/H]}<-1$, no star with $\mbox{[Fe/H]} <-4$ has yet been found. Larger telescopes with multiplexing capabilities are required to significantly improve the number of observed stars, as this apparent lack is likely simply a result of few observed stars to date (often only 1-3 stars per galaxy).

The prospect of finding a surviving first star is truly exciting, but even non-detections allow us to put stringent limits on the low-mass end of the Pop~III IMF. For example, \citet{Hartwig15b} estimate the expected numbers of low-mass Pop~III stars in the Galactic halo based on semi-analytic models of the early star formation history in Milky Way-like galaxies. They conclude that if no metal-free star is ever found in a sample of 4 million Milky Way stars then we can exclude the existence of low-mass Pop~III stars with masses below $0.8\,${\Ms} with a confidence level of 68 per cent. No detection in a sample of 20 million stars would even exclude the existence of these stars with a statistical significance of $>99$ per cent.

Here, we extend this analysis by improving the stellar feedback model and investigating in detail a set of 30 merger trees constructed from the \textit{Caterpillar} project \cite[see][]{caterpillar}. This allows us to predict the likelihood of finding genuine low-mass Pop~III survivors in satellite galaxies of the Milky Way. Our improved model has several advantages over previous works which also model the potential existence of Pop~III stars in the Milky Way and its satellite galaxies. 

These previous studies generally use one of two approaches. The simplest approach \citep[see e.g.][]{tumlinson2006,salvadori2007,debennassuti2014,Hartwig15b,komiya16} is to construct a merger tree for the progenitors of a Milky Way-like galaxy using the extended Press-Schechter formalism \citep{Bond1991,LaceyCole1993}. Starting with the lowest-mass, earliest progenitors and moving forward in time, a simple semi-analytic model is then applied to determine which progenitors are capable of forming Pop~III stars, and how many such stars with masses below $0.8\,${\Ms} form in each progenitor. In this way, the number of such stars present in the Milky Way at the present day can be predicted. A similar method can also be used to gather information on the metallicity distribution of low metallicity Pop~II stars. 

This approach has the disadvantage that it contains no spatial information: we learn nothing about the spatial distribution of low-mass Pop~III stars within the Milky Way, and also learn little about which of its satellites are the best places to search for these stars. A better approach is therefore to draw the merger trees directly from a high resolution N-body simulation of a Milky Way-like galaxy \citep[see e.g.][]{scannapieco2006,salvadori2010,tumlinson2010,Gao10,ishiyama16,Graziani17}. This allows the location of Pop~III stars associated with Milky Way progenitor halos to be followed over time (under the assumption that the stars follow the dark matter), and hence allows one to make predictions for where these stars end up.

\citet{Griffen17} also modeled Pop~III star formation based on the \textit{Caterpillar} merger trees. The main differences to their study lies in our less heuristic modeling of chemical feedback (Section \ref{sect:chem}) and the addition of spatially inhomogeneous ionizing radiation feedback (Section \ref{sect:ion}). Further differences are caused by the strength and impact of the LW background we employed, but these are mere variations in model parameters, rather than in the methods.

However, previous efforts along these lines generally continue to treat the effects of radiative and chemical feedback from ongoing star formation in a position-independent way, i.e.\ the spatial location of a particular progenitor halo is not taken into account when determining whether it is affected by feedback. In addition, these studies typically consider only one or a few N-body simulations (limiting their statistical power), and the simulations themselves often do not have sufficiently high resolution to resolve the full range of halos capable of forming Pop~III stars. The method we present in this paper overcomes all of these important limitations.

\section{NUMERICAL METHOD}
\label{sect:method}
\subsection{Merger Trees \& Pop~III Formation}
Modelling Milky Way-like environments with a semi-analytical, merger tree-based model requires a careful selection of host haloes. While statistical algorithms \citep[e.g.][]{Parkinson2008, Galform} can be used to generate merger trees in the appropriate mass range, they can not fully account for the characteristic local environmental conditions of the Milky Way host such as the distance to other massive haloes. Additionally, the extended Press-Schechter formalism \citep{Bond1991, LaceyCole1993} which constitutes the core of most statistical algorithms, does not describe the substructures of haloes, which is important for this work.

To overcome the challenges with statistical algorithms, we base our model on the high-resolution, dark matter only \textit{Caterpillar} simulation suite first presented in \citet{caterpillar}. This suite contains 30 haloes that are similar in mass to that of the Milky Way host. The haloes were selected from a parent simulation, which has a periodic box of $l \approx 100\,h^{-1}$\,Mpc, $1024^3$ particles (particle mass $\sim 1.22 \times 10^7\,\Ms$), and assumes the Planck 2013 cosmology given by $\Omega_m = 0.32,\ \Omega_\Lambda = 0.68,\ \Omega_b = 0.05,\ n_s = 0.96,\ \sigma_8 = 0.83$ and Hubble constant, $H= 67.11\,\mathrm{km\,s}^{-1}\mathrm{Mpc}^{-1}$ \citep{Planck2013}. Initial conditions were created using \textsc{MUSIC} \citep{music}.

The highest resolution zoom-in region of the target haloes were run at an effective resolution of $(2^{14})^{3}$ particles which corresponds to $2.99 \times 10^4\,\Ms$ per particle. The temporal resolution between each snapshot is $\sim$5 Myr down to $z = 6$ and $\sim$50\,Myr to $z = 0$. Haloes are identified with an improved version of the halo finding algorithm \textsc{Rockstar} \citep{rockstar} using full iterative unbinding (important for haloes on highly radial orbits) and merger trees are constructed using \textsc{Consistent-Trees} \citep{ConsistentTrees}. \textsc{Rockstar} assigns virial masses to haloes, $M_\mathrm{vir}$, using the evolution of the virial relation from \citet{Bryan1998}. At $z = 0$, this definition corresponds to an over-density of $104$ times the critical density of the Universe.

We have additionally modelled one halo at higher resolution with a particle mass of $3.73 \times 10^3\,\Ms$ and have described our convergence study in Appendix \ref{sect:resol}. For more detail of the simulation parameters, see \citet{caterpillar}.

Using merger trees from N-body simulations has the crucial advantage over most statistical implementations that the positions and velocities of the haloes are known. Therefore our model of feedback can account for halo clustering and the potentially complex correlation between merger history and halo position. Not being able to do so has been a major caveat to many previous efforts to study the assembly of the Milky Way with a merger tree-based approach \citep[e.g.][]{tumlinson2006,Hartwig15b,komiya16,Graziani17}.

Our model is based on \citet{Hartwig15b, Hartwig16a, Hartwig16b} and \citet{Magg16}, but major parts of the feedback modelling are improved to exploit the information we gain by resorting to simulated merger trees. Similar to these studies we follow the merger trees beginning from the highest redshifts ($z_\mathrm{max}\approx 30$) to identify the sites of the first star formation. Haloes are assumed to form Pop~III stars when H$_2$ cooling starts to become efficient, i.e. when they reach a mass of
\begin{equation}
 M_\mathrm{crit} = 10^6\,\Ms\,\left(\frac{T_\mathrm{crit}}{1000\,\mathrm{K}}\right)^{1.5}\left(\frac{1+z}{10}\right)^{-1.5},
 \label{eq:m_crit}
\end{equation}
where we adopt a critical virial temperature $T_\mathrm{crit} = 2200\,$K from \citet{hum12}. We will refer to halos above this threshold as minihaloes and to haloes above $T_\mathrm{crit} = 10^4\,$K as atomic cooling haloes. It has previously been shown that the overall Pop~III star formation history is relatively insensitive to small variations in $T_\mathrm{crit}$\ \citep{Magg16}. When a halo reaches this mass threshold it is assumed to convert a fixed fraction $\eta_\mathrm{III}$ of its gas into Population III stars, i.e.
\begin{equation}
 M_{*,\mathrm{III}} = \eta_\mathrm{III} \frac{\Omega_b}{\Omega_m}M_\mathrm{halo},
\end{equation}
where $M_\mathrm{halo}$ is the virial dark matter mass of the halo. We use a Pop~III star formation efficiency of $\eta_\mathrm{III}=0.002$. This parameter is chosen such that our model reproduces the metallicity distribution function of metal-poor stars in the Galactic halo (Hartwig et al. in prep.). The stars are then sampled from a logarithmically flat IMF \citep{Clark11,Greif11b,Dopcke13} in the range $0.6\,\Ms<M<150\,\Ms$. 
Star formation is suppressed in haloes with a mass growth factor, $\Delta M$, during a redshift step $\Delta z$ that is larger than
\begin{equation}
 \frac{\Delta M}{\Delta z} = 3.3 \times 10^6\,\Ms\left(\frac{M_\mathrm{halo}}{10^6\,\Ms}\right)^{3.9},
\end{equation}
with the idea that in these haloes dynamical heating prevents the gas from cooling and collapsing \citep{Yoshida2003}.

The majority of the chemical and radiative feedback in our model originates from metal-enriched stars, and so the Pop~III star formation rate (SFR) is not very sensitive to the upper limit of the underlying IMF (Hartwig et al. in prep.). The lower mass limit of the pristine IMF is uncertain. While subsolar mass fragments are seen to accumulate in simulations at the sites of first star formation, their abundance is uncertain and it is unclear if they merge to form more massive stars \citep{Greif11b,Latif15,Stacy16,Hirano17}. Our Pop~III SFRs are insensitive to the lower mass limit of the IMF, as low mass stars are ineffective feedback agents. Since our primary goal is to identify the sites of surviving low mass Pop~III stars, the lower mass limit is chosen such that it allows for at least some Pop~III stars to survive to $z\sim0$. Allowing for Pop~III stars with masses below 0.6\,\Ms\ would raise the overall number of Pop~III stars, but not change the implications of our results as these stars are too faint to be identified as genuine Pop~III survivors.

\subsection{Chemical feedback}
\label{sect:chem}
Some of the first stars explode as SNe and enrich their environment with metals. These newly-enriched regions are the birthplaces of the second generation, Population II or Pop~II stars. Haloes that are enriched from the outside accumulate metals in dense clumps and form Pop~II stars \citep{bsmith15, Chen16b}. Without properly simulating the mixing process, the exact metallicity of these clumps is difficult to estimate. Additionally it is uncertain at what metallicity the transition from Pop~III to Pop~II star formation occurs \citep{Frebel07,Omukai10,SchneiderR12,Dopcke13}. Therefore, we only distinguish haloes as pristine or enriched. Each halo that underwent star formation is assigned a radius to which it enriches its environment. A halo is enriched if it or one of its progenitors is inside the enrichment radius of a neighbouring star-forming halo. These enriched haloes are prevented from forming Pop~III stars, but will form Pop~II stars contributing to subsequent enrichment and feedback. As small amounts of metals do not significantly influence the early stages of the collapse of a halo \citep{Jappsen07,Glover14}, we apply the same mass thresholds as for Pop~III-forming haloes to these externally enriched Pop~II-forming haloes.

To model delay between Pop~III formation and the enrichment, we interpolate between the lifetimes of non-rotating metal-free stars provided by \citet{Marigo2001}. Pop~III stars between 11--40\,\Ms\ explode as core collapse SNe \citep{hw02} with an explosion energy of $10^{51}$\,erg. We model the expansion of the shell as
\begin{equation}
 R(t) = v_\mathrm{III}t,
\end{equation}
with $v_\mathrm{III}=10\,$km\,s$^{-1}$, estimated from the expansion of the supernova remnant in \citet[][Figure 1]{bsmith15}. We approximate the expansion of the remnant as ballistic because \citet{bsmith15} find that the deceleration of the shell is roughly compensated by additional momentum from subsequent Pop~III SNe. Similar to our Pop~III star formation model, enriched haloes are assumed to convert a fixed fraction $\eta_\mathrm{II}$ of their gas mass into Pop~II stars
\begin{equation}
 M_{*,\mathrm{II}} = \eta_\mathrm{II} \frac{\Omega_b}{\Omega_m}M_\mathrm{halo},
 \label{eq:MII}
\end{equation}
where $\eta_\mathrm{II} = 0.02$. Whenever the halo mass increases, a fixed fraction of the accreted mass is turned into Pop~II stars, such that Equation \eqref{eq:MII} remains valid. In the host halo of the exploding SN remnant, Pop~II star formation is delayed because the gas is heated and evacuated from the halo. Estimates of how long the re-collapse of this heated gas takes vary between a few $\sim10$\,Myr to several $\sim100$\,Myr \citep[e.g.][]{Greif10,wet13a,bsmith15,Jeon14,Jeon2017}. We prevent Pop~II star formation for a recovery time $t_\mathrm{recov}=100\,\mathrm{Myr}$ to account for this delay.

The outflow mass, M$_{\rm out}$, from Pop II galaxies is parametrized as a function of the stellar mass using a broken power law. Our approach is motivated by \citet{DallaVecchia13}, \citet{Hayward17}, and \citet{Muratov17}. For low mass galaxies, i.e.\ $M_{*,\mathrm{II}} < 2\times 10^6 \Ms$ we assume that the outflow mass is 20 times that of the stellar mass. For galaxies larger than this, we use a flatter power-law, as described below: 
\begin{equation}
 M_\mathrm{out} = \begin{cases}
                  20 M_{*,\mathrm{II}},& \mathrm{if}  M_{*,\mathrm{II}} < 2\times 10^6\,\Ms\\
                  10^{6.3} M_{*,\mathrm{II}}^{0.2} & \mathrm{otherwise.}
                  \end{cases}
 \end{equation}
We assume that the outflows are launched at
\begin{equation}
 R_\mathrm{launch} = 0.142\,\mathrm{kpc} \left(\frac{M_\mathrm{halo}}{10^8\,\Ms}\right)^{1/3}\left(\frac{1+z}{10}\right)^{-1},
\end{equation}
which is 10 per cent of the virial radius. The shell expands as a momentum-driven snowplough i.e.\ it expands with constant momentum, and sweeps up the intergalactic medium (IGM). Therefore, if the mass density of baryons $\rho_\mathrm{b}$ in the IGM is assumed to be the cosmological average i.e.
\begin{equation}
 \rho_\mathrm{b} = \Omega_\mathrm{b} \frac{3 H_0^2}{8\pi G}(1+z)^3,
\end{equation}
the velocity of the metal-enriched shell that has expanded to radius $R_\mathrm{enr}$ can be written as
\begin{equation}
 v_\mathrm{II} = v_\mathrm{out}\frac{M_\mathrm{out}}{M_\mathrm{out}+\frac{4}{3}\pi \rho_b\left(R_\mathrm{enr}^3-R_\mathrm{vir}^3\right)}.
\end{equation}
We use an outflow velocity of $v_\mathrm{out} = 110$\,km\,s$^{-1}$ \citep{agarw12}. The shell expansion is stalled when its speed equals the current speed of sound $c_\mathrm{s}$. Since the shells mostly expand into the ionized IGM we assume a value of $c_\mathrm{s}= 10$\,km\,s$^{-1}$.
\subsection{Radiative feedback}
\label{sect:ion}
The dominant forms of radiative feedback from the first and later generations of stars are ionizing and Lyman-Werner (LW) radiation. We treat ionizing radiation based on a simple shell model similar to the chemical feedback. Under the assumption of spherical symmetry and a homogeneous IGM we compute the radius of the ionized region around every Pop~II-forming halo. If a halo forms Pop~II stars with a SFR of $\dot{M}_*$ the rate of ionizing photon production is
\begin{equation}
 \Ni = \frac{\dot{M}_*}{m_p} a_i f_\mathrm{esc,i},
\end{equation}
where $a_i=4000$ is the number of ionizing photons per stellar baryon \citep{GreifBromm06}, $f_\mathrm{esc,i}$ is the ionizing photon escape fraction, which we assume to be 0.1 \citep{wise12a,Paardekooper2013} and $m_p$ is the proton mass. For this approximation we neglect the lifetimes of the stars and assume all ionizing photons are produced instantaneously when the stars form, which is a reasonable assumption because most of the ionizing photons are produced by young massive stars. These photons contribute either to the expansion of the ionized region, or to keeping it ionized against recombination. In other words the rate at which a halo produces ionizing photons can be written as
\begin{equation}
 \Ni = Vn^2C\alpha + \dot{V} n,
 \label{eq:ion_rate}
\end{equation}
where $V$ and $\dot{V}$ denote the volume of the ionized region and its time-derivative, respectively, $n = 0.75 \rho_\mathrm{b} / m_p$ is the current average hydrogen number density, $C$ is the clumping factor and $\alpha=2.6\times10^{-13}\,\mathrm{cm}^3\,\mathrm{s}^{-1}$ \citep{ISM_phys} is the case B recombination rate coefficient of atomic hydrogen in a $10^4\,$K IGM. We assume $C=3$ \citep{Robertson13}, but note that our model is insensitive to the exact value, as the recombination only acts as a minor correction to the expansion of the ionized regions. Computing the evolution of the region in terms of the volume, rather than in terms of the radius allows us to use an implicit Euler method for integrating Equation \eqref{eq:ion_rate}, which guarantees numerical stability. If the ionized volume at time-step $i$ is $V_i$, the volume at the next time step, i.e. with an increment of $\Delta t$ will be
\begin{equation}
  V_{i+1} = V_{i} +\Delta t \dot{V}(V_{i+1}) = V_{i} + \frac{\Ni \Delta t}{n} - n\alpha C V_{i+1}\Delta t,
\end{equation}
which can explicitly be solved for $V_{i+1}$:
\begin{equation}
  V_{i+1} = \left(V_i +\frac{\Ni \Delta t}{n}\right)\left(1+\Delta t n\alpha C\right)^{-1}.
\end{equation}
The implicit integration prevents us from overestimating the size of the ionized regions, especially in the case of strong starbursts or large time-steps. As the ionized regions are small compared to cosmological scales, we do not consider the effect of cosmological expansion and redshifting of ionizing photons in our model. 

Being located in such an ionized region, haloes might still form Pop~III stars \citep[see e.g.][]{Finlator17, Visbal17}. Thus we allow pristine haloes within such an ionized region to form Pop~III stars if their virial temperature (Equation \ref{eq:m_crit}) is above $10^4$\,K, which corresponds to a circular velocity of $v_\mathrm{circ}=10\,\mathrm{km\,s}^{-1}$ \citep{Dijkstra2004}. \citet{Visbal17} find that this approximation is only valid in regions where the photoionizing flux is lower than $\sim6.7\times10^5$\,photons\,s$^{-1}$\,cm$^{-2}$. However, the region where the flux is high is mostly enriched with metals anyway in our model. As an illustration we consider typical halo properties at $z=7$, where star formation in atomic cooling haloes peaks in our model. \citet{Visbal17} cite an ionizing flux of $6.7\times10^5$\,photons\,s$^{-1}$\,cm$^{-2}$ at 50\,kpc distance from a halo with $M_\mathrm{vir}=7\times10^{11}\Ms$. This flux marks the boundary at which the critical mass for haloes grows beyond the $T_\mathrm{vir}=10^4$\,K threshold. Our Milky Way progenitors are typically an order of magnitude less massive, which results in the same flux being reached at 16\,kpc distance from the halo, if we assume the flux scales linearly with the halo mass. At the same time our model typically predicts a metal-enriched bubble with a radius $R_\mathrm{enr}\sim$22\,kpc and an ionized region with a radius $R_\mathrm{ion}\sim 250\,$kpc. Therefore, it is reasonable to assume that most, if not all, pristine, atomic cooling haloes inside an ionized region will form Pop~III stars.

LW feedback is implemented as a uniform background and adds an additional mass threshold for Pop~III formation. At a given redshift, for a LW {flux} $F_{21}$ (in units of $10^{-21}$\,erg\,s$^{-1}$\,cm$^{-2}$\,Hz$^{-1}$), we use an additional minimum mass threshold for Pop~III star formation from \citet{Oshea08}
\begin{equation}
 M_\mathrm{LW} = 5\times 10^5\,\Ms + 34.8\times 10^5\,\Ms F_{21}^{0.47}.
\end{equation}
We implement a global LW background that evolves as
\begin{equation}
 F_{21} = 4\pi\times 10^{-\frac{z-z_0}{5}},
\end{equation}
where $z_0=10$ \citep{GreifBromm06,Ahn09,agarw12}. We do not consider LW radiation from close by haloes, as the local fluctuations of the LW flux affect a relatively small fraction of all the haloes at a given redshift \citep{Dijkstra08, Ahn09, agarw12}. Additionally, for Pop~III-forming haloes, the escape fractions of LW photons tend to be much smaller in the near field feedback case \citep{Schauer2015}. The LW escape fractions of Pop~II-forming regions depends on various factors such as the mass of the halo, the IMF, and the star formation history \citep{Schauer17a}. Capturing this complex interplay of parameters is beyond the scope of this project.

\section{Results}
\label{sect:res}
\subsection{SFRs and general simulation properties}
\begin{figure}
 \includegraphics[width=0.98\linewidth]{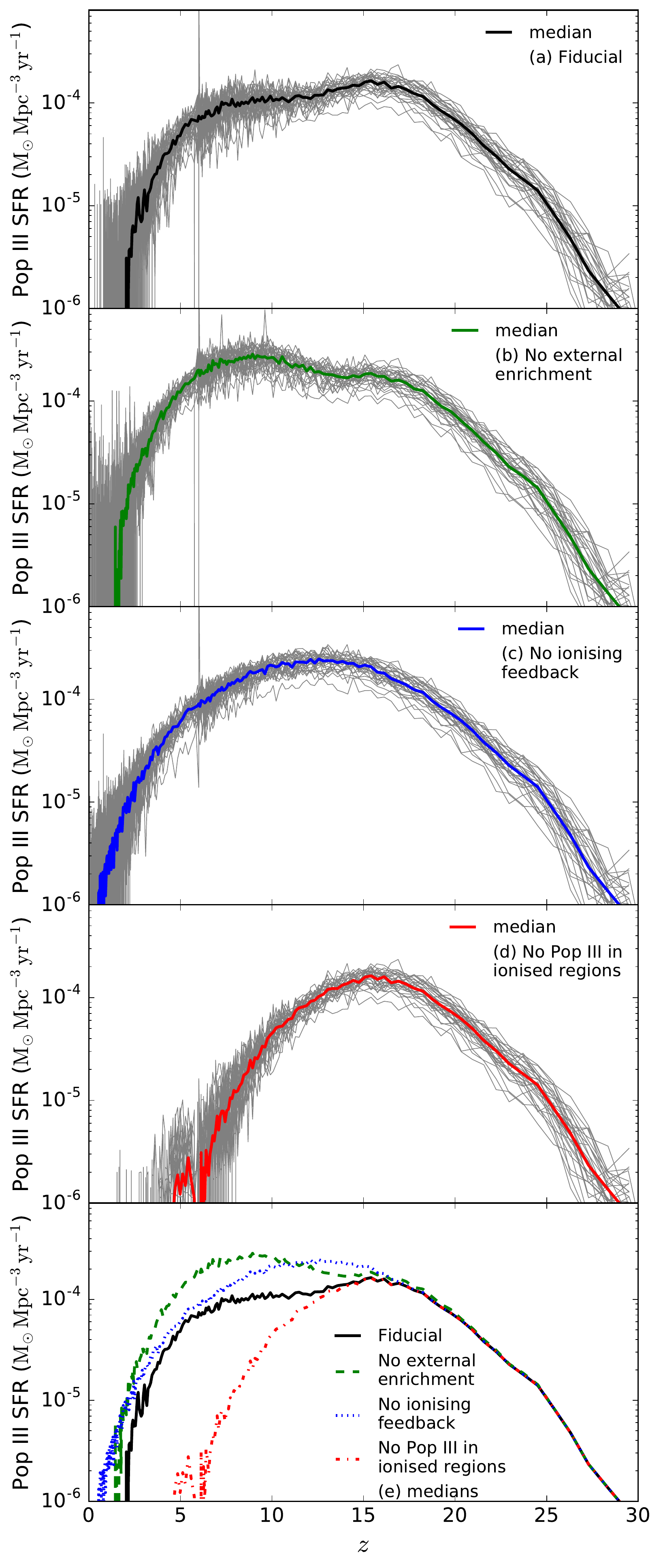}
 \caption{\label{fig:SFRs} Comoving Pop~III SFR densities of the 30 \textit{Caterpillar} boxes (grey) in units of solar masses per comoving Mpc$^3$ per year and their medians. We show four different cases: our fiducial model with full physics (a), deactivated external metal enrichment (b), deactivated feedback from ionizing radiation (c) and for an extreme ionization case (d), in which Pop~III star formation is not allowed inside the ionized regions, not even in the most massive haloes. Panel (e) shows the medians from all four cases in one plot, for easier comparison. The fiducial SFR has its peak at $z\approx 17$ and starts to fall again at $z\la 5$.}
\end{figure}
In Figure \ref{fig:SFRs} we show the comoving Pop~III SFR density of our fiducial model and three cases with reduced physics to better understand the contributions of different processes. In the full-physics fiducial case the Pop~III SFRs peak at $z\approx 17$, stay roughly flat until $z\approx 5$, and decrease afterwards. The maximum is caused by the onset of reionization around the most massive haloes, which shuts off star formation in their vicinity. External metal enrichment suppresses the Pop~III SFR by about a factor of three at $z\sim 6$ where it has the biggest impact. It has no significant effect above $z\approx15$. Such a small effect of external metal enrichment was also found by \citet{Visbal17b} and \citet{Jaacks17}. The decrease of the Pop~III SFR at the lowest redshift is caused by a combination of all feedback effects, as well as the increasing halo mass corresponding to a $T_\mathrm{vir} = 10^4\,\mathrm{K}$ star formation threshold (see Equation \ref{eq:m_crit}). The vertical grey lines around $z\sim 6$ are numerical artefacts, caused by the changing time-step of the merger trees, i.e. $\Delta t \sim 5$\,Myr to $z$ = 6 and $\Delta t \sim$50\,Myr thereafter.

\begin{figure}
 \includegraphics[width=0.98\linewidth]{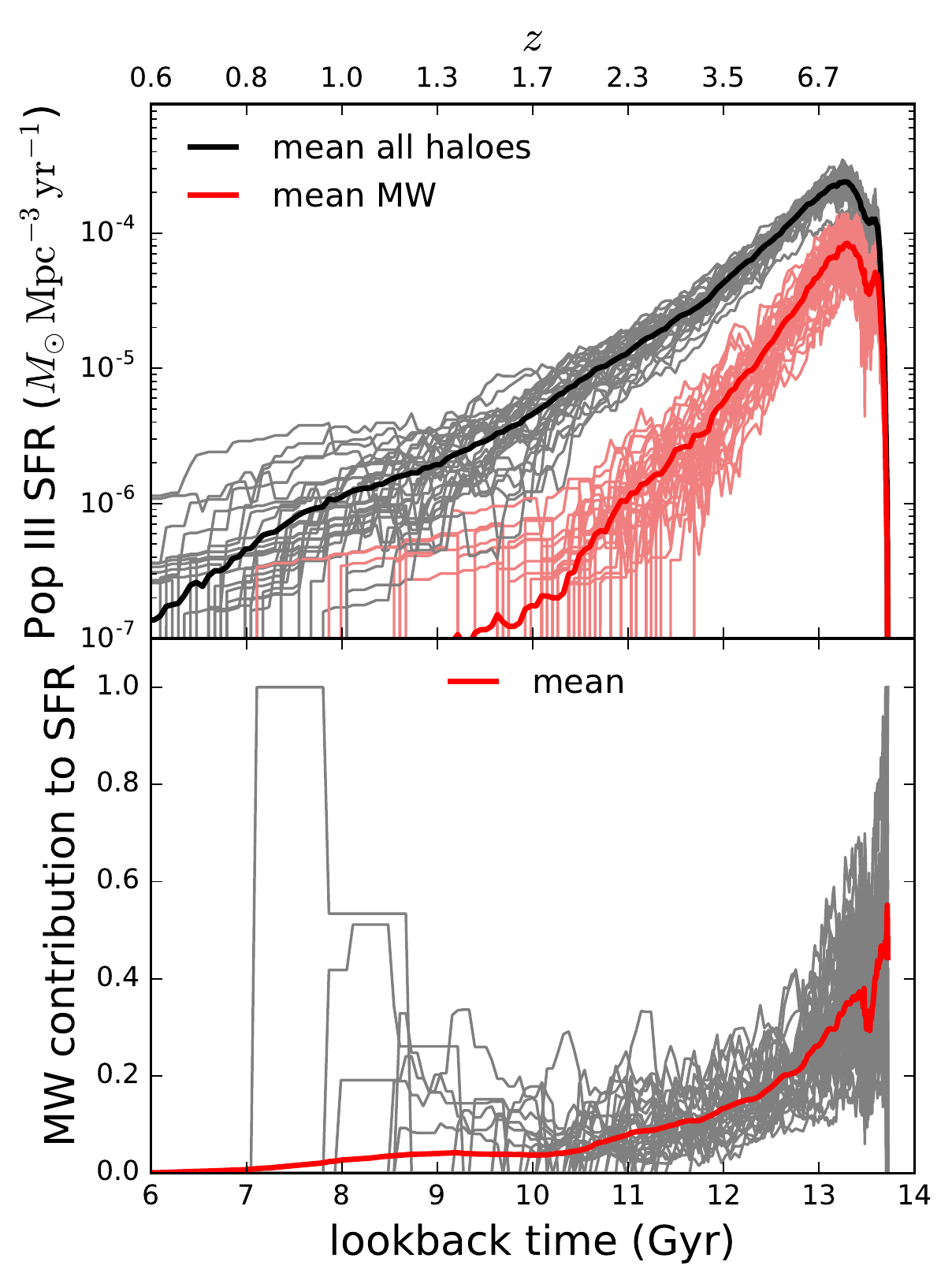}
 \caption{\label{fig:SFR_frac}Upper panel: Comoving Pop~III SFR densities inside the 30 \textit{Caterpillar} boxes (grey) and their mean (black). In light red (all boxes) and red (mean) we show the part of these SFRs that comes from the progenitors of the main haloes and their subhaloes. Lower panel: fraction of star formation that takes place in the progenitors of the main halo and its satellites. At low redshifts SFRs are dominated by single star-bursts and they have been averaged over $\Delta z=0.2$ for clarity. Pop~III formation in the Milky Way and satellite progenitors declines much earlier than in the rest of the simulated regions.}
\end{figure}

Our model predicts high Pop~III formation rates after reionization is complete. While late Pop~III formation around $z\sim 7.6$ is also reported in \citet{Xu16}, it is unclear whether it would continue down to lower redshifts. To further analyse star formation at low redshift, we plot the Pop~III SFR and the fraction of star formation that occurs in the progenitors of Milky Way haloes and their subhaloes in Figure \ref{fig:SFR_frac}. At lookback times of $\le$10\,Gyr, only a few per cent of the Pop~III formation happens in the Milky Way and satellite progenitors. Therefore, the Pop~III formation at the lowest redshifts occurs mostly in the outskirts of the simulated regions and has only a small effect on our predictions regarding surviving Pop~III stars in the Milky Way and its satellites. We might miss additional feedback from outside the modelled regions, which could suppress the late Pop~III formation in these outskirts.

\begin{figure}
 \includegraphics[width=0.98\linewidth]{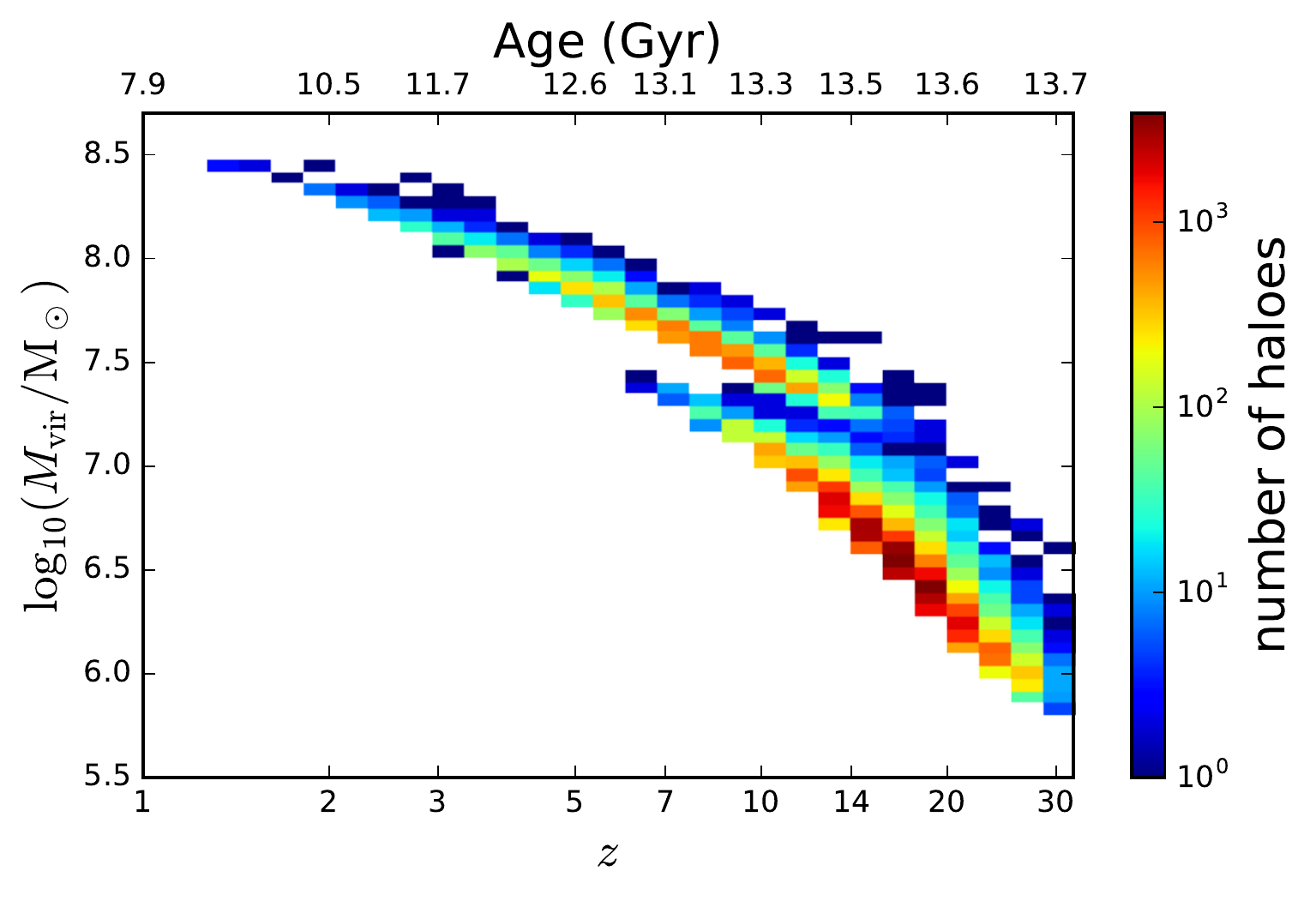}
 \caption{\label{fig:SF_bim}Two dimensional histogram of Pop~III forming haloes as function of redshift and virial mass in all progenitors of Milky Way haloes and their subhaloes. At high redshifts ($z\ga 17$), Pop~III formation occurs predominantly in H$_2$-cooled haloes. At redshift $z\la 6$ the region is fully ionized and Pop~III formation takes place only in atomic cooling haloes.}
\end{figure}
To illustrate the two modes of Pop~III formation, i.e. in H$_2$-cooled minihaloes and in atomic cooling haloes, we plot the number of Pop~III forming haloes as function of their redshift and virial mass in Figure \ref{fig:SF_bim}. Only the progenitors of Milky Way haloes and their subhaloes are taken into account from now on. Some minihaloes grow by a factor of a few larger than the required critical mass before they form stars. This occurs if a halo grows rapidly when crossing the Pop~III formation mass threshold, and its star formation is therefore suppressed by dynamical heating. In the redshift range $6<z<20$ Pop~III formation occurs both in atomic cooling and in H$_2$ cooling haloes. Above $z\approx17$ almost no Pop~III formation occurs in atomic cooling haloes. At $z\approx6$ the region in which the Milky Way progenitors are located is fully ionized and no star formation can take place in H$_2$ cooling haloes. Half of Pop~III stars form ionized regions at $z\approx 12$, which is roughly consistent with the redshift of reionization $z = 11.52$ given by \citet{Planck2013}. However, it is not clear whether we should expect the ionization history to match the Planck results, as we are not modelling a representative region of the Universe, but instead focus on small regions that will form systems like our Local Group.

\begin{figure}
 \includegraphics[width=0.99\linewidth]{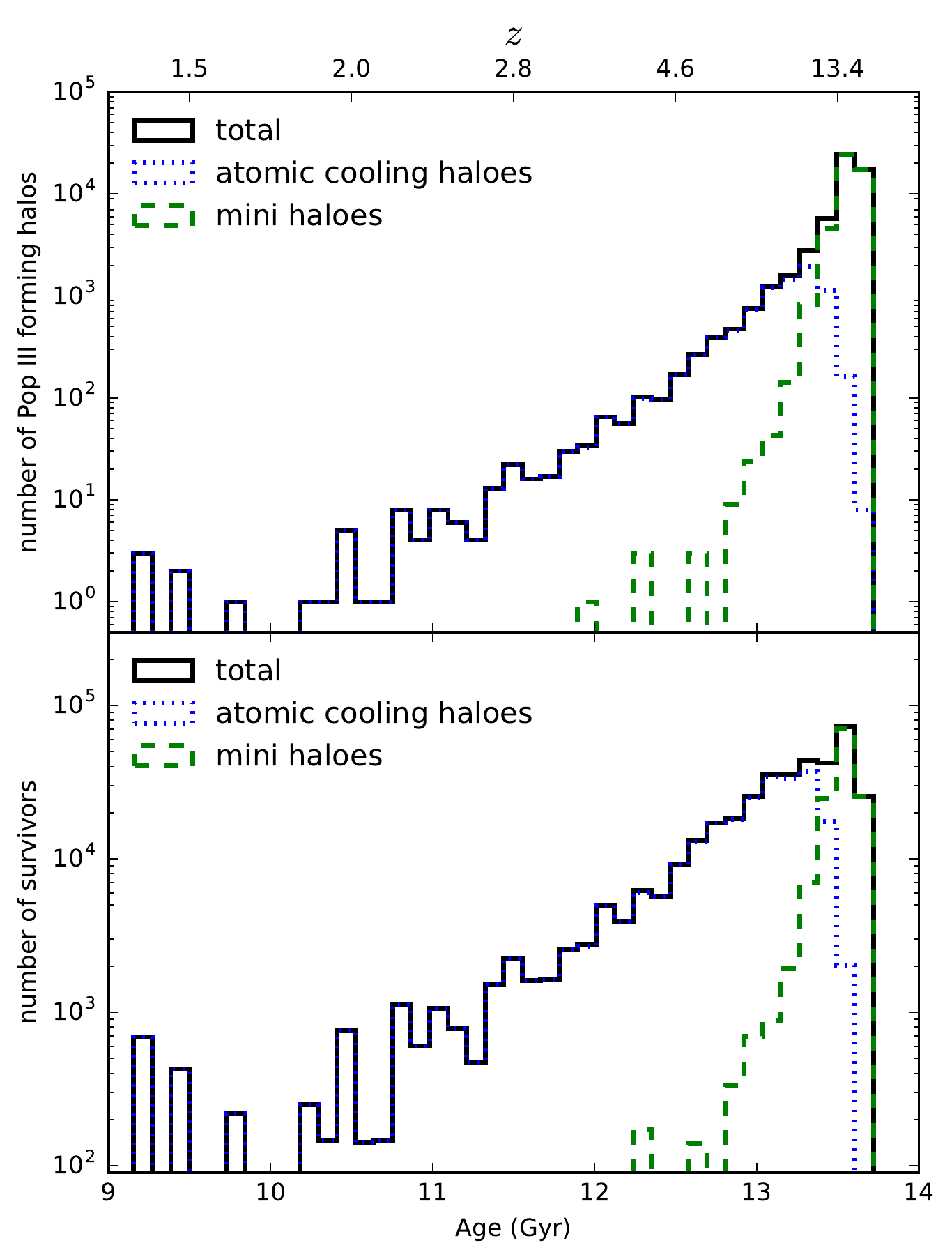}
 \caption{\label{fig:age}Upper panel: ages of Pop~III star bursts. Lower panel: distribution of ages of Pop~III survivors. Most survivors have ages larger than $12\,$Gyr. However, there are rare Pop~III formation events much later. These are caused by massive haloes that stay pristine to low redshifts. We plot histograms for total numbers (black, solid), only Pop~III formation in atomic cooling haloes (blue, dotted) and only in minihaloes (green, dashed) separately. Atomic cooling haloes dominate at all but the earliest times.}
\end{figure}

Figure \ref{fig:age} shows the distribution of ages of surviving Pop~III stars in the Milky Way and its satellites. 70 per cent of Pop~III stars and 96 per cent of Pop~III starbursts have ages larger than 13\,Gyr. Pop~III host haloes after reionization are much rarer, but also more massive than their high-redshift counterparts, thus they produce a larger number of Pop~III survivors per halo. 95 per cent of Pop~III survivors and 99 per cent of Pop~III starbursts are older than 12\,Gyr. Pop~III survivors younger than 10\,Gyr are extremely rare. Therefore, it is highly unlikely that Pop~III survivors more massive than $M_*\sim0.8\,\Ms$ exist in the Milky Way or its satellites.

\begin{figure}
 \includegraphics[width=0.98\linewidth]{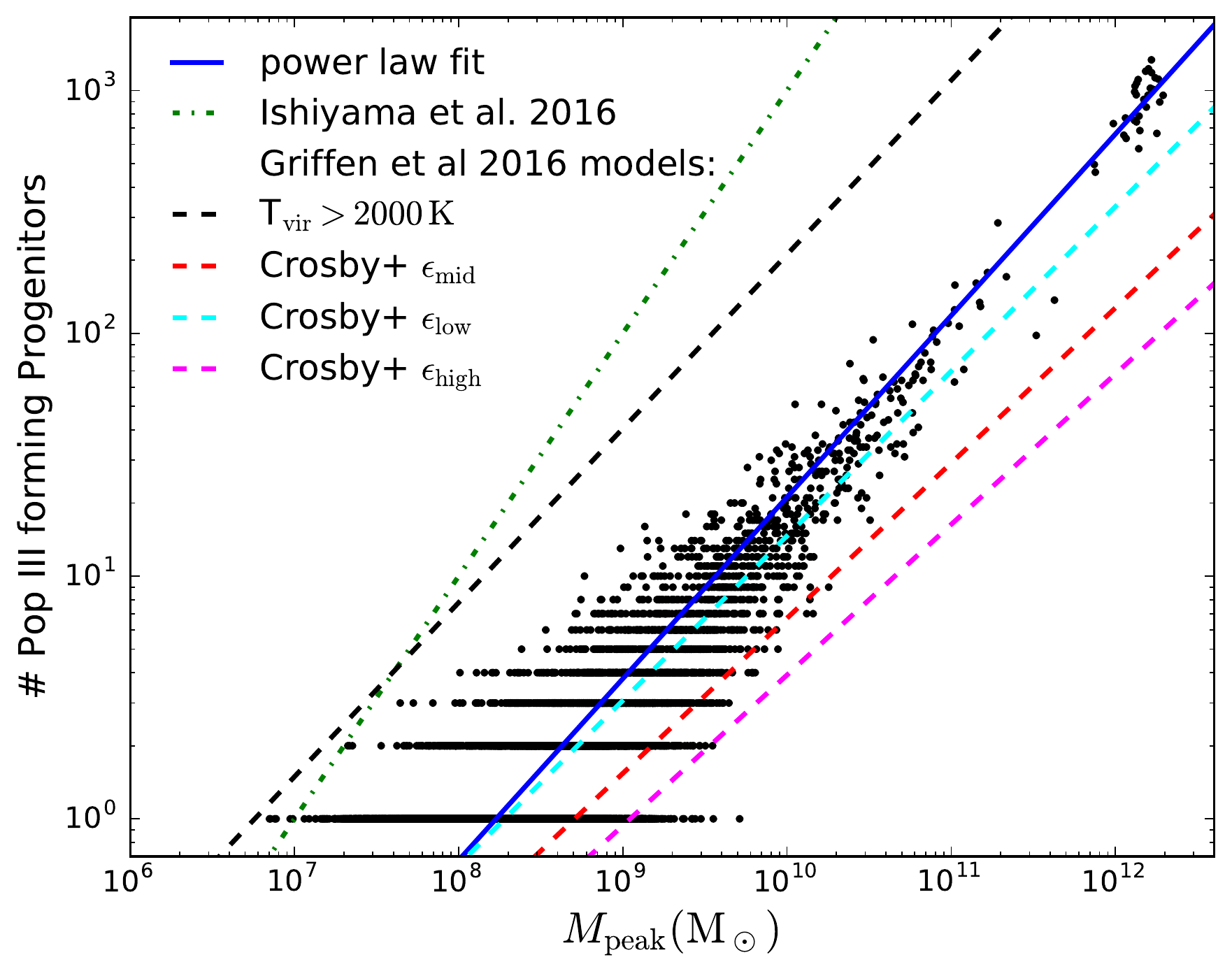}
 \caption{\label{fig:prog} The number of progenitors that form Pop~III stars as a function of the peak dark matter mass of the Milky Way-like haloes and their subhaloes. The solid line is a power-law fit to our data. We show similar power-law fits (dashed) to four different cases of LW feedback from \citet{Griffen17} for comparison: no LW feedback (black), weak feedback (cyan), medium feedback (red) and strong feedback (magenta). The Pop~III-forming progenitor numbers from \citet[dashed-dotted]{ishiyama16} are noticeably higher than ours.}
\end{figure}

To compare to other semi-analytical models of Pop~III formation, we show the number of Pop~III forming progenitors as a function of the peak dark matter mass in Figure \ref{fig:prog}. The power-law fit shows that the average number of Pop~III forming progenitors is
\begin{equation}
 N_\mathrm{Pop\,III\,prog} = 7.1\times10^{-7}\left( \frac{M_\mathrm{peak}}{\Ms}\right)^{0.75}.
\end{equation}
The model of \citet{ishiyama16} predicts a notably higher number of Pop~III forming progenitors, especially at high masses. This is primarily due to their much lower halo mass threshold for Pop~III formation. The fit to our fiducial model shows a similar slope as the four \citet{Griffen17} models, and it is lower than their no LW feedback and but higher than their weak LW feedback case. This is again mainly caused by the different mass threshold for Pop~III formation: their LW feedback model is based on critical masses from \citet{Crosby13}, which exceeds both our $T_\mathrm{vir}=2200\,$K and atomic cooling mass thresholds \citep[compare our Figure \ref{fig:SF_bim} to Figure 1 in][]{Griffen17}.

\subsection{Pop~III survivors}
\begin{figure}
 \includegraphics[width=0.98\linewidth]{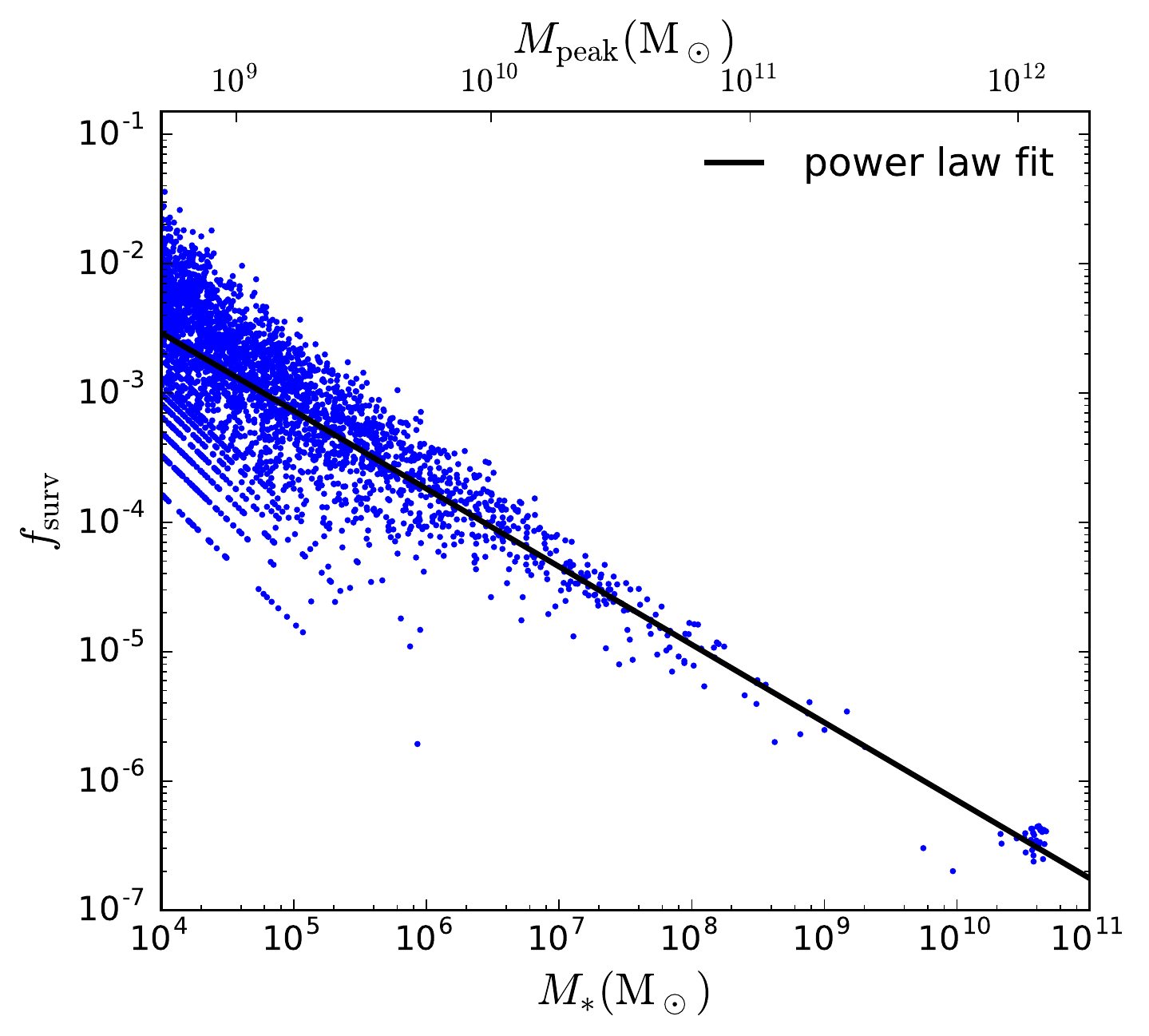}
 \caption{\label{fig:surv_law} Fraction of surviving Pop~III stars as a function of stellar mass for all Milky Way-like haloes and their subhaloes. The cluster of haloes around $M_*\approx 4 \times 10^{10}\,\Ms$ are the 30 Milky Way-like haloes. The stellar masses have been determined with the \citet{GarrisonKimmel} abundance matching scheme. The survivor fraction generally increases towards lower stellar masses.}
\end{figure}

We identify subhaloes of the 30 main haloes as their satellite galaxies. At $z=0$, we use the \citet{GarrisonKimmel} abundance matching scheme to assign stellar masses to these simulated Milky Way satellites. A halo that reaches a peak mass of $M_\mathrm{peak}$ during its evolution is assigned a stellar mass $M_*$, where 
\begin{equation}
 \log\left(\frac{M_*}{\Ms}\right) = \log(\epsilon M_1)+f\left(\log\left(\frac{M_\mathrm{peak}}{M_1}\right)\right)-f(0),
\end{equation}
with $M_1= 3.27\times10^{11}\,\Ms$, $\epsilon=0.0167$ and
\begin{equation}
 f(x)= -\log\left(10^{-1.92 x}+1\right)+\frac{3.5 \left(\log(1+\exp(x)\right)^{0.32}}{1+\exp\left(10^{-x}\right)}.
\end{equation}
This scheme is similar to the abundance matching from \citet[][Equation 3]{Behroozi2013}, but has a steeper power-law slope, causing the stellar mass to evolve as
\begin{equation}
 M_*\approx 3\times10^6\,\Ms \left(\frac{M_\mathrm{peak}}{10^{10}\,\Ms}\right)^{1.92}
\end{equation}
at $M_* \la 10^8\,\Ms$. Figure \ref{fig:surv_law} shows the fraction of all stars that are Pop~III survivors ($f_\mathrm{surv}$) as a function of stellar mass. The 30 Milky Way host haloes are seen as a cluster around $M_*\approx 4 \times 10^{10}\,\Ms$. In these haloes, we find between 1800 and 5200 Pop~III survivors, which is well below the upper limit for the abundance of Pop~III survivors derived in \citet{Hartwig15b}. A detailed analysis of the likelihood of finding Pop~III survivors in the Milky Way requires a careful consideration of how the survivors are distributed in the Milky Way \citep{ishiyama16,komiya16}. As our model only provides halo positions and not positions of the stars within a given halo, this analysis is beyond the scope of this project.

In Figure \ref{fig:surv_law} we see that satellite galaxies with low stellar masses have a higher fraction of Pop~III survivors. Fitting a power law to the Pop~III survivor fraction results in
\begin{equation}
 f_\mathrm{surv} = 0.37\left(\frac{M_*}{\Ms}\right)^{-0.62}.
\end{equation}
The slope of the fit implies that, while the overall fraction of Pop~III survivors decreases with increasing stellar mass, the average number of Pop~III survivors ($N_\mathrm{surv}\propto f_\mathrm{surv}M_*\propto M_*^{0.4}$) increases. However, at $M_*\la 10^6\,\Ms$ the scatter in the Pop~III fraction is larger than an order of magnitude. Additionally, at masses lower than $M_*\la 10^{5}\,\Ms$ the abundance matching relation becomes very uncertain as this mass regime is incompletely sampled \citep{GarrisonKimmel17,Bullock17}. Still, the trend to find higher Pop~III fractions at lower masses is evident.

\begin{figure*}
 \includegraphics[width=1.9\columnwidth]{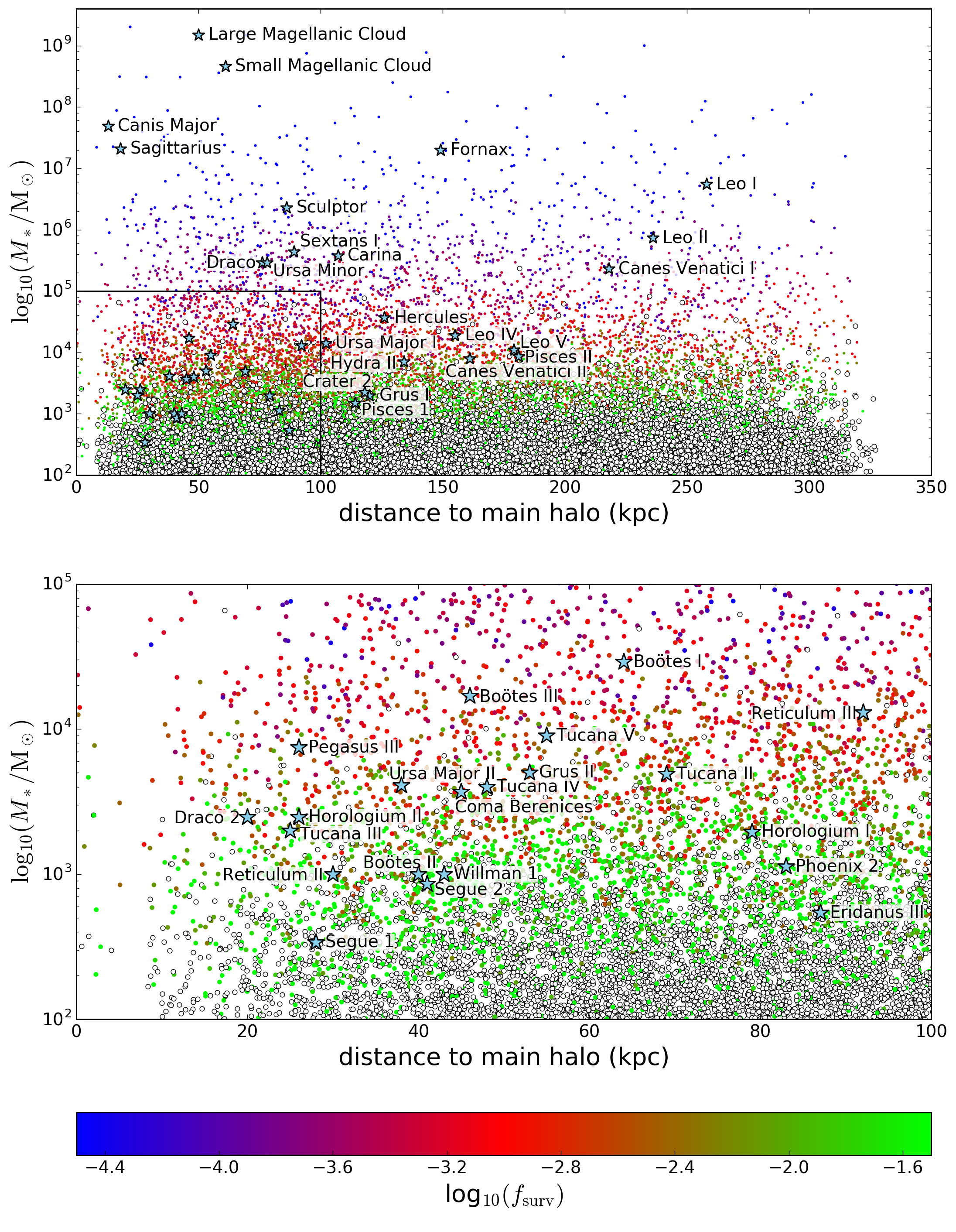}
 \caption{\label{fig:satellites} Coloured circles show simulated Milky Way satellites from all 30 \textit{Caterpillar} realisations as a function of distance from the Milky Way centre and stellar mass. Colour-coded is the expected fraction of stars that are Pop~III survivors. Open circles represent modelled subhaloes without Pop~III survivors. Observed Milky Way satellites, shown as blue stars, are from \citet{McConnachie2012} and \citet{Sales17}. A list of the masses and distances used in this plot is provided in Table \ref{tab:props}. Less massive satellites have a larger Pop~III fraction, but satellites below $M_*\approx 10^3\Ms$ often have not formed Pop~III stars at all.}
\end{figure*}

Figure \ref{fig:satellites} shows the fraction of Pop~III survivors of all modelled Milky Way satellites as a function of stellar mass and distance to the centre of the main halo. As seen already in Figure \ref{fig:surv_law} the fraction of Pop~III survivors increases towards lower halo masses. A large fraction of satellites below $M_* \approx 10^3\,\Ms$ have not undergone Pop~III formation or do not contain any surviving pristine stars. Haloes that do not host Pop~III survivors become rare above stellar masses of a few thousand solar masses. We do not find any notable trends with the distance from the Milky Way-like haloes.

\subsection{Expected number of stars on the red giant branch}

\begin{figure}
 \includegraphics[width=0.99\linewidth]{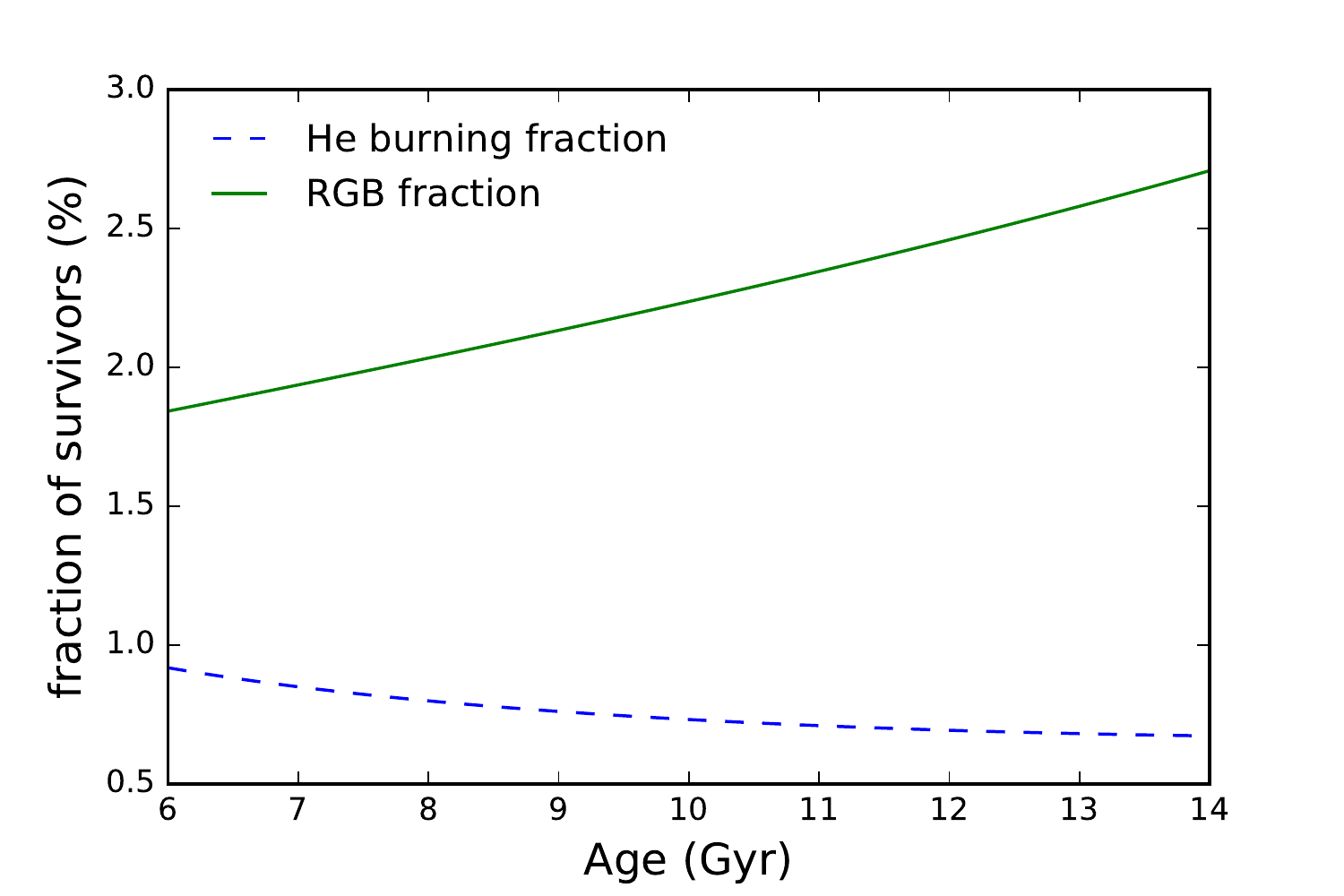}
 \caption{\label{fig:giant_fraction} Fraction of Pop~III survivors that are in their red giant branch phase (green solid line) or HB phase (blue dashed line) as a function of the age of the stellar population. HB fractions are computed using interpolated lifetimes from \citet{Marigo2001}, while the red giant branch fraction is based on an interpolation of the $Z=0.001$ models from \citet{Bertelli08}}
\end{figure}

A satellite containing a large fraction of Pop~III stars does not necessarily imply that actual Pop~III survivor stars are readily identifiable. The challenge in finding any lies in the following. 
Only low mass Pop~III stars would have sufficiently long lifetimes to survive until today. If any of these were ever formed, the most massive of them would have around 0.8\,\Ms and now, after $\sim$13\,Gyr, would be located on the red giant branch. Less massive ones would be on the main sequence and the subgiant branch.

To increase the chance of confirming a candidate as an actual low-mass Pop~III star, a high resolution spectrum with high signal-to-noise would be required because the lack of iron and other metal lines would need to be established. Poor data quality might not allow one to distinguish between a true Pop~III star and an extremely metal-poor star with weak metal lines. The best candidates are thus luminous red giants, which are comparably brighter and also have intrinsically stronger metal lines at a given metal abundance. Both aspects would make it easier to identify a star as metal-poor as opposed to metal-free. A surviving Pop~III star would then be one that only has extremely low upper limits on all metal abundances.

In addition, when searching for Pop~III stars in dwarf satellite galaxies, observations are naturally restricted to just the very brightest red giants in each dwarf given their large distances out into the Galactic halo. Stars with magnitudes down to $V\sim19$ are barely observable at present, with exposure times taking 6-10\,h per star on an 8m-class telescope. Fainter stars are not accessible with current telescopes and high-resolution spectrographs, which severely limits the detailed exploration of dwarf galaxies. The high-resolution spectrograph for the Extremely Large Telescope \citep[ELT-HIRES,][]{ELT-HIRES} will make many fainter stars accessible for such observations and may thus be ideally suited for finding surviving Pop~III stars.

Accordingly, we determine the fraction of Pop~III survivor stars that are expected to be in the red giant branch and also in the horizontal branch phase of their evolution. We use the main sequence lifetimes, $\tau_\mathrm{H}$, of the stars and the duration of the the two evolutionary phases, $\tau_\mathrm{P}$. For a specific age, $\tau_\mathrm{a}$, we calculate the two masses at which stars satisfy $\tau_\mathrm{a} =\tau_\mathrm{H}$ and $\tau_\mathrm{a}=\tau_\mathrm{H}+\tau_\mathrm{P}$. 
The fraction of survivor stars in the desired evolutionary stage is then just the fraction of survivors that fall in this mass range. This is obtained by integrating our Pop~III IMF. Figure \ref{fig:giant_fraction} shows the fractions of Pop~III survivors in the red giant branch and horizontal branch stages. 

For the horizontal branch, we use H and He burning times from \citet{Marigo2001}. Due to uncertainty of the duration of the red giant branch phase in the \citet{Marigo2001} models, we use the $Z=0.001$ models from \citet{Bertelli08} to constrain the red giant branch star fraction. The total lifetimes of subsolar mass stars in \citet{Marigo2001} and in the $Z=0.001$ models of \citet{Bertelli08} are very similar. The red giant branch fraction is around 2.5 per cent for the relevant stellar ages (compare Figure \ref{fig:age}). We adopt this red giant branch fraction for all further analysis. The horizontal branch population is $<1$ per cent. In addition, these stars have slightly lower luminosities than stars on the tip of the red giant branch. We conclude that these stars are likely of minor relevance to the identification of surviving Pop~III stars in Milky Way satellites.

\begin{figure*}
 \includegraphics[width=1.9\columnwidth]{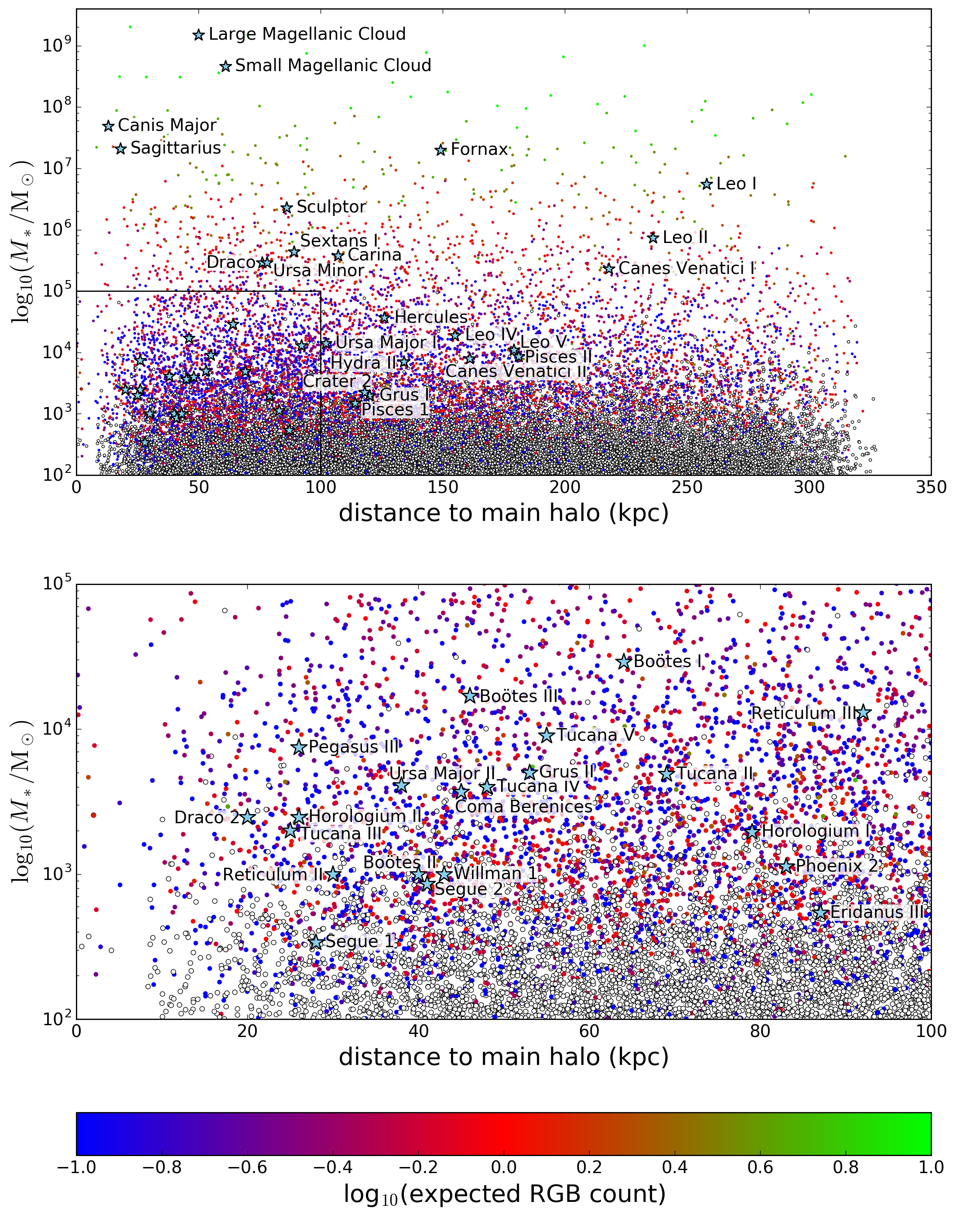}
 \caption{\label{fig:RGB_scatter} Same as Figure \ref{fig:satellites} but with expected number of red giant branch stars colour-coded instead of the Pop~III fraction. While lower mass satellites have a higher Pop~III fraction, the expected number of Pop~III red giant branch stars is often too low to make detections of a Pop~III survivor likely.}
\end{figure*}

We show the expected number of Pop~III red giant branch stars in our simulated haloes in Figure \ref{fig:RGB_scatter}. Since only a small fraction of the Pop~III survivors are in that phase, the averages shown are subject to additional Poisson noise. As compared to haloes with $M_* \approx 10^4\,\Ms$, haloes with larger masses tend to host more Pop~III red giant branch stars, even though their Pop~III fraction is smaller. Satellites above $M_*\approx10^5$\,\Ms\ host one or more Pop~III red giant branch stars on average. Overall, we find that in the entire mass range covered by Milky Way satellites, some systems are expected to host one or even a few Pop~III red giant branch stars.

\renewcommand{\arraystretch}{1.2}
\begin{table*}
\begin{tabular}{lrrlllrrc}
\hline
Name & Stellar Mass & Distance & $n_\mathrm{surv}$ & \# RGB stars & $f_{\mathrm{surv}}$ & $P_0$ & $N_\mathrm{av}$ & Source \\
 & \Ms & $\mathrm{kpc}$ &  &  & $\mathrm{\%}$ & $\mathrm{\%}$ &  &   \\ \hline
Milky Way & $6 \times 10^{10} $ & 0 & $3756_{2811}^{4854}$ & $93.9_{70.3}^{121}$ & $(\mathrm{1.7}_\mathrm{1.4}^\mathrm{1.9})\times 10^{-5}$ & 0 & 30 & c \\
Large Magellanic Cloud & $1.5\times 10^9$ & 50 & $1257$ & $31$ & $1.4 \times 10^{-4}$ & 0 & 1 & a \\
Small Magellanic Cloud & $4.6\times 10^8$ & 61 & $509$ & $13$ & $2.3\times 10^{-4}$ & 0 & 1 & a \\
Canis Major & $4.9\times 10^7$ & 13 & $230$ & $5.8$ & $4.3\times 10^{-4}$ & 0.25 & 1 & a \\
Sagittarius & $2.1\times 10^7$ & 18 & $69$ & $1.7$ & $5.0\times 10^{-4}$ & 18 & 1 & a \\
Fornax & $2\times 10^7$ & 149 & $179_{118}^{201}$ & $4.5_{3}^{5}$ & $0.0013_{0.001}^{0.0018}$ & 2.8 & 9 & a \\
Leo I & $5.5\times 10^6$ & 258 & $91_{60}^{178}$ & $2.3_{1.5}^{4.5}$ & $0.0032_{0.0019}^{0.0053}$ & 12 & 11 & a \\
Sculptor & $2.3\times 10^6$ & 86 & $69_{38}^{130}$ & $1.7_{0.95}^{3.3}$ & $0.0055_{0.0026}^{0.008}$ & 21 & 13 & a \\
Leo II & $7.4\times 10^5$ & 236 & $33_{19}^{72}$ & $0.83_{0.48}^{1.8}$ & $0.0079_{0.0039}^{0.014}$ & 41 & 23 & a \\
Sextans I & $4.4\times 10^5$ & 89 & $30_{11}^{61}$ & $0.76_{0.29}^{1.5}$ & $0.013_{0.0038}^{0.026}$ & 45 & 36 & a \\
Carina & $3.8\times 10^5$ & 107 & $30_{16}^{87}$ & $0.76_{0.4}^{2.2}$ & $0.014_{0.0075}^{0.035}$ & 42 & 42 & a \\
Draco & $2.9\times 10^5$ & 76 & $26_{14}^{49}$ & $0.65_{0.36}^{1.2}$ & $0.015_{0.0073}^{0.027}$ & 50 & 46 & a \\
Ursa Minor & $2.9\times 10^5$ & 78 & $26_{14}^{48}$ & $0.66_{0.36}^{1.2}$ & $0.015_{0.0073}^{0.027}$ & 50 & 46 & a \\
Canes Venatici I & $2.3\times 10^5$ & 218 & $24_{11}^{56}$ & $0.6_{0.28}^{1.4}$ & $0.017_{0.0078}^{0.041}$ & 52 & 67 & a \\
Hercules & $3.7\times 10^4$ & 126 & $11_{3}^{34}$ & $0.28_{0.075}^{0.86}$ & $0.052_{0.016}^{0.15}$ & 69 & 125 & a \\
Bo\"otes I & $2.9\times 10^4$ & 64 & $10_{4}^{32}$ & $0.26_{0.1}^{0.82}$ & $0.058_{0.023}^{0.18}$ & 71 & 70 & a \\
Leo IV & $1.9\times 10^4$ & 155 & $10_{2}^{30}$ & $0.25_{0.058}^{0.77}$ & $0.082_{0.021}^{0.27}$ & 71 & 178 & a \\
Bo\"otes III & $1.7\times 10^4$ & 46 & $9_{5}^{25}$ & $0.23_{0.12}^{0.63}$ & $0.086_{0.044}^{0.25}$ & 73 & 55 & a \\
Ursa Major I & $1.4\times 10^4$ & 102 & $9_{2}^{29}$ & $0.23_{0.05}^{0.73}$ & $0.11_{0.029}^{0.35}$ & 73 & 171 & a \\
Reticulum III & $1.3\times 10^4$ & 92 & $8_{2}^{28}$ & $0.2_{0.05}^{0.7}$ & $0.11_{0.031}^{0.35}$ & 75 & 176 & b \\
Leo V & $1.1\times 10^4$ & 179 & $10_{2}^{36}$ & $0.25_{0.05}^{0.9}$ & $0.15_{0.031}^{0.55}$ & 70 & 244 & a \\
Tucana V & $9\times 10^3$ & 55 & $8_{2}^{28}$ & $0.21_{0.055}^{0.72}$ & $0.18_{0.048}^{0.5}$ & 74 & 96 & b \\
Pisces II & $8.6\times 10^3$ & 181 & $9_{2}^{41}$ & $0.23_{0.05}^{1}$ & $0.17_{0.039}^{0.76}$ & 69 & 252 & a \\
Canes Venatici II & $7.9\times 10^3$ & 161 & $10_{2}^{41}$ & $0.25_{0.05}^{1}$ & $0.19_{0.05}^{0.89}$ & 68 & 250 & a \\
Pegasus III & $7.5\times 10^3$ & 26 & $7_{3}^{27}$ & $0.18_{0.075}^{0.69}$ & $0.14_{0.058}^{0.64}$ & 75 & 28 & b \\
Hydra II & $7.1\times 10^3$ & 134 & $9_{3}^{41}$ & $0.23_{0.075}^{1}$ & $0.21_{0.06}^{0.98}$ & 68 & 237 & b \\
Grus II & $5\times 10^3$ & 53 & $8_{1}^{26}$ & $0.2_{0.025}^{0.66}$ & $0.26_{0.039}^{0.91}$ & 75 & 122 & b \\
Tucana II & $4.9\times 10^3$ & 69 & $9_{2}^{31}$ & $0.23_{0.05}^{0.79}$ & $0.29_{0.055}^{1.1}$ & 73 & 197 & b \\
Ursa Major II & $4.1\times 10^3$ & 38 & $7_{2}^{31}$ & $0.18_{0.05}^{0.8}$ & $0.31_{0.083}^{1.1}$ & 74 & 89 & a \\
Tucana IV & $4\times 10^3$ & 48 & $5_{1}^{26}$ & $0.12_{0.025}^{0.65}$ & $0.23_{0.036}^{1.1}$ & 78 & 116 & b \\
Coma Berenices & $3.7\times 10^3$ & 45 & $6_{2}^{25}$ & $0.15_{0.05}^{0.65}$ & $0.26_{0.08}^{1.2}$ & 77 & 120 & a \\
Draco 2 & $2.5\times 10^3$ & 20.0 & $8_{1}^{34}$ & $0.2_{0.025}^{0.87}$ & $0.63_{0.067}^{2.4}$ & 70 & 34 & b \\
Horologium II & $2.5\times 10^3$ & 26 & $7_{1}^{33}$ & $0.19_{0.044}^{0.84}$ & $0.55_{0.1}^{2.3}$ & 72 & 62 & b \\
Crater 2 & $2.2\times 10^3$ & 118 & $6_{0}^{38}$ & $0.15_{0.0}^{0.96}$ & $0.45_{0}^{3}$ & 74 & 399 & b \\
Tucana III & $2\times 10^3$ & 25 & $6_{0}^{33}$ & $0.15_{0.015}^{0.83}$ & $0.45_{0.038}^{2.4}$ & 77 & 61 & b \\
Horologium I & $2\times 10^3$ & 79 & $6_{0}^{31}$ & $0.15_{0.0}^{0.78}$ & $0.51_{0.0}^{2.6}$ & 76 & 299 & b \\
Grus I & $2\times 10^3$ & 120 & $6_{0}^{40}$ & $0.15_{0.0}^{1}$ & $0.51_{0.0}^{3.3}$ & 74 & 441 & b \\
Pisces 1 & $1.5\times 10^3$ & 114 & $5_{0}^{43}$ & $0.12_{0.0}^{1.1}$ & $0.62_{0.0}^{4.9}$ & 73 & 492 & b \\
Phoenix 2 & $1.1\times 10^3$ & 83 & $4_{0}^{33}$ & $0.1_{0.0}^{0.84}$ & $0.62_{0.0}^{4.7}$ & 76 & 423 & b \\
Reticulum II & $1\times 10^3$ & 30 & $4_{0}^{28}$ & $0.1_{0.0}^{0.7}$ & $0.57_{0.0}^{4.2}$ & 81 & 108 & b \\
Willman 1 & $1\times 10^3$ & 43 & $5_{0}^{28}$ & $0.12_{0.0}^{0.7}$ & $0.74_{0.0}^{4.9}$ & 79 & 209 & a \\
Bo\"otes II & $1\times 10^3$ & 40 & $4_{0}^{26}$ & $0.1_{0.0}^{0.66}$ & $0.63_{0.0}^{4.7}$ & 80 & 193 & a \\
Segue 2 & $8.6\times 10^2$ & 41 & $3_{0}^{25}$ & $0.075_{0.0}^{0.63}$ & $0.5_{0.0}^{5.0}$ & 82 & 206 & a \\
Eridanus III & $5.4\times 10^2$ & 87 & $0_{0}^{23}$ & $0.0_{0.0}^{0.58}$ & $0.0_{0.0}^{6.4}$ & 87 & 625 & b \\
Segue 1 & $3.4\times 10^2$ & 28 & $0_{0}^{3}$ & $0.0_{0.0}^{0.075}$ & $0.0_{0.0}^{1.5}$ & 94 & 115 & a \\ \hline
\end{tabular}
\renewcommand{\arraystretch}{1.0}
\caption{\label{tab:props} Inferred properties of dwarf galaxies. Columns from left to right: Name of the satellite, stellar mass in \Ms, estimated distance to centre of the Milky Way in kpc, total number of Pop~III survivors (median, 16th and 84th percentile), expected number of red giant Pop~III survivors (median, 16th and 84th percentile), Pop~III fraction (median, 16th and 84th percentile), probability to host no Pop~III red giant branch stars, number of simulated haloes contributing estimated properties of the dwarfs, sources a: \citet{McConnachie2012}, b: \citet{Sales17}, c: \citet{Licquia15}. Medians and percentiles are computed from all haloes within a factor of 1.3 in stellar mass and distance from the Milky Way. For some of the most massive satellites there are no haloes in this range of parameters and we give the values for the simulated halo that is closest to the satellite's mass and distance. For the Milky Way haloes we use all 30 main haloes.}
\end{table*}

\subsection{Satellite properties and implications for observations}

We now estimate the number of Pop~III survivor stars (red giant branch stars and total number) and the overall Pop~III fraction that can be expected to be present in each Milky Way satellite dwarf galaxy. This is shown in Table \ref{tab:props}. The number counts are based on average values from our modelled subhaloes that have similar properties to the actual dwarf galaxies. Specifically, we consider all subhaloes that have a stellar mass and a distance to their host halo which are within a factor of 1.3 of a Milky Way satellite to be similar to the satellite. For observed satellites for which this parameter region does not yield any analogues in our simulation, we adopt values from the closest simulated subhalo. 

Table \ref{tab:props} illustrates the basic trade-off necessary when searching for Pop~III survivor stars in dwarf galaxies. In lower mass satellites, Pop~III survivors make up a larger fraction of the stellar mass which should in principle make finding them easier. In practice, however, the actual expected number of observable Pop~III red giant branch stars in such systems is very small and might even be zero. On the other hand, in satellites large enough to likely contain observable Pop~III stars, the Pop~III fraction is very small given the large number of metal-enriched stars that have formed in these systems. This means that many more stars need to be observed in a dwarf galaxy to actually find a Pop~III survivor star or to at least derive a meaningful upper limit on the number of potential survivors.

We also provide the likelihood for each dwarf galaxy to host no Pop~III red giant branch stars. This probability is derived by randomly choosing a halo that is similar to the observed satellite galaxy (as defined above) from our simulations. Each Pop~III survivor star in that halo is set to be a red giant branch star with a 2.5 per cent probability of existence. We repeat this process 10000 times for each satellite to ensure well-sampled statistics. This means that for a satellite galaxy with only one simulated equivalent we sample the same subhalo 10000 times, while for a satellite galaxy with many simulated counterparts all of them are sampled uniformly. The three most massive satellite galaxies and the Milky Way always contain red giant Pop~III stars in our model. Dwarf galaxies above $M_* \approx 10^5\,\Ms$ have a probability of containing one or more Pop~III red giant branch stars that is about 50 per cent or larger. However, satellite galaxies less massive than $M_* \approx 10^5\,\Ms$ host Pop~III red giant branch stars only with a probability of $\sim 70$ per cent or less. Therefore, while they offer the greatest chance of finding genuine Pop~III survivors, observations of individual galaxies of this type are ill-suited for constraining the Pop~III IMF by non-detections.

Recently, some dwarf galaxies have been targeted for deep follow up, either with low- and medium-resolution spectroscopy \citep[e.g.][Chiti et al. in prep.]{Kirby15} or with narrow-band photometry such as the \textit{Pristine} photometric survey \citep{Starkenburg17}. For Bo\"otes\,I and Hercules, the two most massive satellites targeted in \textit{Pristine}, we estimate an average number of Pop~III red giant branch stars to be about 0.3. There is a 51 per cent probability that at least one of these two galaxies contains at least one Pop~III red giant branch star. Thus, there is a good chance of finding a metal-free star in these galaxies if enough stars are observed. This means that, in the end, the search for a low-mass Pop~III star comes down to identifying dwarf galaxies for which large numbers of stars can be chemically characterized with high precision to weed out all stars with any metal lines present until a Pop~III star is found.

\begin{figure}
 \includegraphics[width=0.99\linewidth]{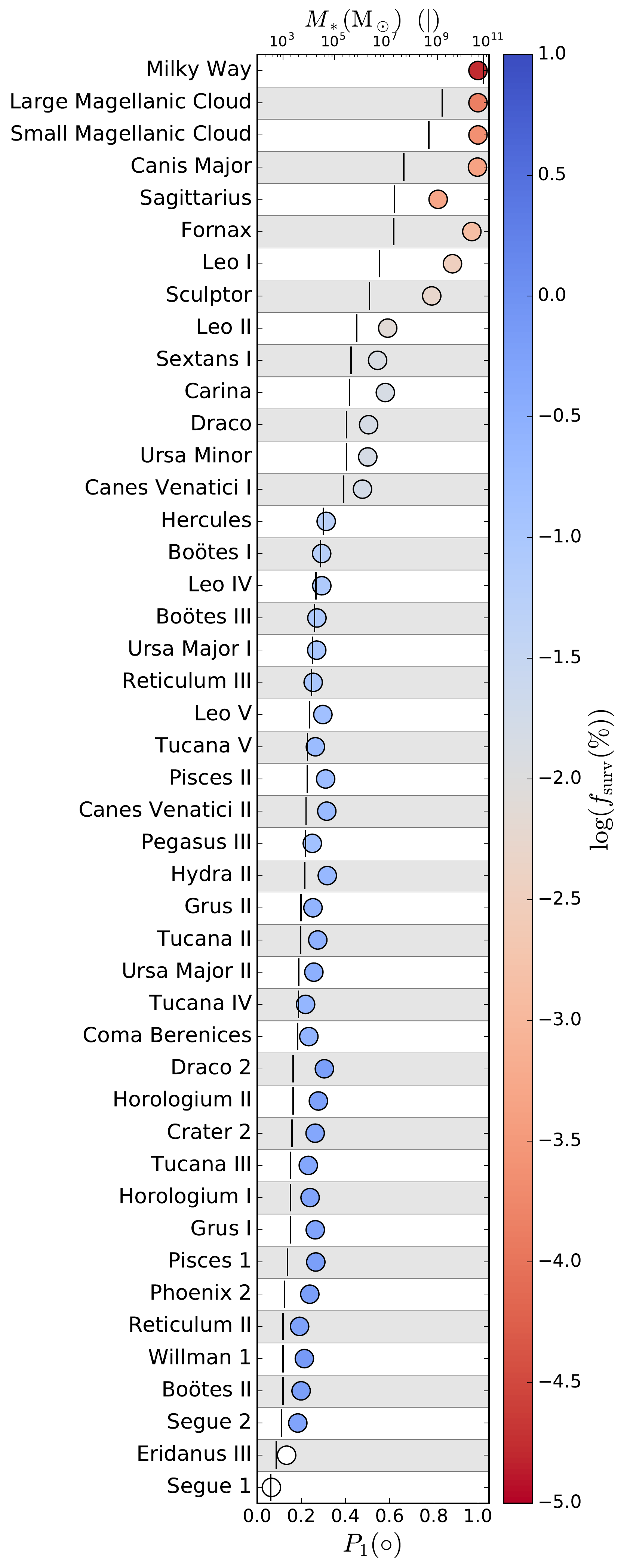}
 \caption{\label{fig:sat_col} Properties of Milky Way dwarf satellite galaxies. The vertical lines indicate the stellar mass as inferred from observations (upper x-axis). The filled circles show for each satellite the probability that it contains at least one Pop~III red giant branch star (lower x-axis). The circles are colour-coded by the median fraction of Pop~III survivors.}
\end{figure}
We visualise the dwarf galaxy properties presented in Table \ref{tab:props} in Figure \ref{fig:sat_col}. Due to their high survivor fraction, we suggest targeting several low mass, nearby satellite galaxies such as Reticulum 2, Willman 1 or Draco 2 for optimal chances of finding Pop~III survivors. If, however, the goal of observations is to constrain the Pop~III IMF by making use of non-detections, we propose to focus on larger satellites (e.g.\ Canis Major or the Sagittarius dSph). These galaxies are very likely to contain Pop~III survivors and thus allow more solid constraints on the Pop~III IMF if none are found. However, such an analysis requires a large sample size of observed stars, which may remain infeasible for some time.

\section{Caveats}
\label{sect:caveat}
Due to the simplicity of our model, we had to make assumptions that impact our results. Most importantly, our model is based on a Pop~III IMF that reaches below 0.8\,\Ms, i.e. we assume that there are Pop~III stars that survive for 10\,Gyr and longer. It is not certain whether these pristine surviving stars actually exist. Some recent simulations show the existence of subsolar mass fragments in primordial star forming regions \citep{Stacy10,Greif11b,Stacy14, Stacy16}. It is not clear how common these low-mass fragments are and whether they evolve to become low-mass metal free stars that survive until today, or whether they merge to become larger stars with shorter lifetimes. Also these low-mass fragments do not occur in all recent simulations \citep{Hirano14, Hosokawa16}. However, this could be due to differences in the resolution or numerical approach of the different calculations. The goal of this study is to predict where Pop~III survivors should be looked for under the assumption that they exist. At the same time, our model can be used to constrain the lower mass limit of the Pop~III IMF \citep[similar to][]{Hartwig15b}, if attempts to find metal-free stars in Milky Way satellites remain fruitless.

If low mass surviving Pop~III stars still exist in the Milky Way or its satellites, they might have been enriched with metals they accreted from the interstellar medium in the course of their lifetimes, which could prevent them from being accurately identified as genuine Pop~III stars. However, this effect is predicted to be only of small significance for most stars \citep{Frebel09, Johnson11}. Metals are most efficiently accreted when a star passes through a dense metal-rich clouds, which is expected to happen only for a small fraction ($\la 0.1$) of metal-free stars \citep{Johnson11}. Stellar winds and radiation fields would further reduce the amount of metals accreted or lead to particular abundance patterns \citep{Johnson15} such that stars might be identified as Pop~III survivors despite having accreted metals over their lifetime. As the metallicity of the interstellar medium and abundances of dense gas are generally lower in Milky Way satellites than in the main Galaxy, we expect accretion of metals during the life of the stars to be even less significant in the satellites. Assuming metal enriched gas is accumulated via Bondi-Hoyle accretion, \citet{Komiya15} and \citet{Shen17} find significant pollution of Pop~III survivors. The pollution could be enhanced by the relative motion of gas and stars being generally smaller in low mass galaxies. However, recently \citet{Tanaka17} demonstrated that the development of a magnetosphere around Pop~III survivors can significantly reduce the efficiency of the accretion of the ambient medium, suggesting that the pristine observational signatures of these stars will be better preserved.

Being based on merger trees, our model will only accurately trace positions of Pop~III survivors if these survivors actually remain in their original host haloes. Pop~III survivors could be removed from small haloes by tidal stripping, without completely disrupting these haloes. \citet{ishiyama16} employ particle tagging techniques for tracing Pop~III survivors, which not only allows one to trace the positions of Pop~III stars within the haloes hosting them, but also results in more reliable positions of Pop~III survivors. They find that Pop~III stars are not significantly affected by tidal stripping before their host halo is disrupted (T. Ishiyama, priv.\ comm.).

Small mass Pop~III stars could be ejected from their host haloes by close stellar encounters \citep{Greif11b, Stacy13} or if their binary companions explode as SNe \citep{komiya16}. This effect reduces the number of Pop~III survivors that can be found in each minihalo. In this case our assumption that Pop~III survivors stay associated with the halo they formed in constitutes an upper limit.

The stellar masses we derived for the Milky Way subhaloes are based on an abundance matching scheme from \citet{Behroozi2013} and \citet{GarrisonKimmel}, which crucially depends on the stellar masses being a monotonous function of the peak halo mass and on the completeness of detections of dwarf galaxies. Both assumptions are uncertain at low stellar masses. \citet{GarrisonKimmel17} have shown that including scatter in the relation between peak mass and stellar mass leads to a more realistic population of Milky Way satellites. This scatter becomes as large as 1\,dex at low halo masses \citep{Munshi17}. We do not account for such scatter as it is unclear how it should correlate with the number of Pop~III stars hosted by the halo. While it could increase or decrease the scatter in the Pop~III fraction or the Pop~III red giant branch count at a given stellar mass, the overall trends in the behaviour should stay unaffected. Below $M_* \sim 10^5\,\Ms$ the census of dwarf galaxies is believed to be incomplete. Any result given for less massive satellites should therefore be treated with care.

In addition, by adopting an abundance matching scheme to relate halo masses to stellar masses, we are implicitly making two further assumptions. First, we are assuming that the process responsible for the fact that the number of known Milky Way satellite galaxies is much smaller than the predicted number of dark matter subhaloes does not affect the subhalo mass function. If this mismatch is ultimately due to the effects of stellar feedback and/or cosmological reionization, then this assumption is reasonable. However, other proposed solutions to the problem exist that work by modifying the properties of the dark matter itself, and consequently changing the form of the subhalo mass function (e.g.\ warm dark matter). These models are reviewed in \citet{Bullock17}, but they are outside the scope of our current study. Second, we are assuming that the Milky Way halo does not have an unusually low number of dark matter subhaloes, i.e.\ that it is a typical example of a halo of this mass, and not some kind of extreme outlier. This is a reasonable assumption in the absence of evidence to the contrary, but it remains to be verified.

Finally, as seen in Figure \ref{fig:prog}, the number of Pop~III-forming progenitors of a halo depends on the required mass for a halo to collapse. Using stronger LW feedback \citep[such as in][]{Griffen17} can reduce the number of Pop~III-forming progenitors by as much as an order of magnitude. A similar increase in the critical mass can be caused by supersonic baryonic streaming motions \citep[][Schauer et al. in prep.]{Greif11a, Maio11, Stacy11, Schauer17b}. On the other hand, accounting for X-ray feedback alongside the LW feedback could partially mitigate its effects \citep{glover2003,machacek2003,ricotti2016}. Large high resolution simulations are needed for finding the critical mass at which a halo can collapse without suffering either from the effects of low resolution or from small number statistics.

\section{Summary and Conclusions}
\label{sect:sum}
By modelling early star formation in spatially resolved merger trees from the \textit{Caterpillar} simulations, we estimate the numbers of surviving Pop~III stars in the Milky Way and its satellites. We find of the order of ten thousand Pop~III survivors in the Milky Way. This number is consistent with not having found any Pop~III survivors until today \citep{Hartwig15b}. Our model is based on the assumption that there are surviving metal-free stars. If no Pop~III survivors are found in future surveys, our model can be used to evaluate the statistical significance of these non-detections.

We identify the most promising Milky Way satellites for searching for surviving Pop~III stars. Lower mass Milky Way satellites typically contain a larger fraction of Pop~III survivors. The two reasons for this trend are that:
\begin{itemize}
  \item the slope of the correlation between peak halo mass and the number of Pop~III-forming progenitors is sub-linear, i.e.\ lower mass satellites have on average more Pop~III-forming progenitors per unit dark matter mass and
  \item the abundance matching slope is steeper than linear, meaning that low mass haloes generally host disproportionally few stars.
\end{itemize}
In particular, we determine the median Pop~III fraction in low mass Milky Way satellites such as Reticulum II to be $10^4$ times as large as the Pop~III fraction of the Milky Way. However, even if these satellite galaxies have a comparatively large fraction of Pop~III stars, the total number of stars is low. In the Milky Way satellites only Pop~III stars in their red giant branch phase are bright enough to allow one to take the high resolution, high signal-to-noise spectra that are needed to put meaningful limits on their metal abundance. When only considering stars that are currently in their red giant branch stage, the expected number of detectable Pop~III survivors decreases further: for our IMF only about 2.5 per cent of the surviving Pop~III stars are red giant branch stars. Thus, when carefully selecting targets and using multi-fibre spectrographs \citep[e.g. G-CLEF][]{G-CLEF14} that enable one to observe several metal-poor red giant branch stars simultaneously, more massive satellites i.e.\ those that are likely to have at least one metal-free red giant branch star, could prove ideal targets.

\section*{Acknowledgements}
We thank the anonymous referee for helping us to significantly improve this paper. We also thank Massimo Ricotti, Naoki Yoshida, Anna Schauer, Tomoaki Ishiyama, and Britton Smith for sharing many helpful thoughts and comments during the preparation of this manuscript. We further thank Volker Bromm for stimulating discussions. The authors were supported by the European Research Council under the European Community's Seventh Framework Programme (FP7/2007 - 2013) via the ERC Advanced Grant `STARLIGHT: Formation of the First Stars' under the project number 339177 (RSK, SCOG and BA) and via the ERC Grant `BLACK' under the project number 614199 (TH). SCOG and RSK also acknowledge funding from the Deutsche Forschungsgemeinschaft via SFB 881 ``The Milky Way System'' (subprojects B1, B2, and B8) and SPP 1573 ``Physics of the Interstellar Medium'' (grant numbers KL 1358/18.1, KL 1358/19.2 and GL 668/2-1). A.F. is supported by NSF CAREER grant AST-1255160. Computations were carried out on the compute cluster of the Astrophysics Division which was built with support from the Kavli Investment Fund, administered by the MIT Kavli Institute for Astrophysics and Space Research.


\bibliographystyle{mnras}
\bibliography{lit}
\appendix
\section{Resolution study}
\label{sect:resol}
\begin{figure}
 \includegraphics[width=0.99\linewidth]{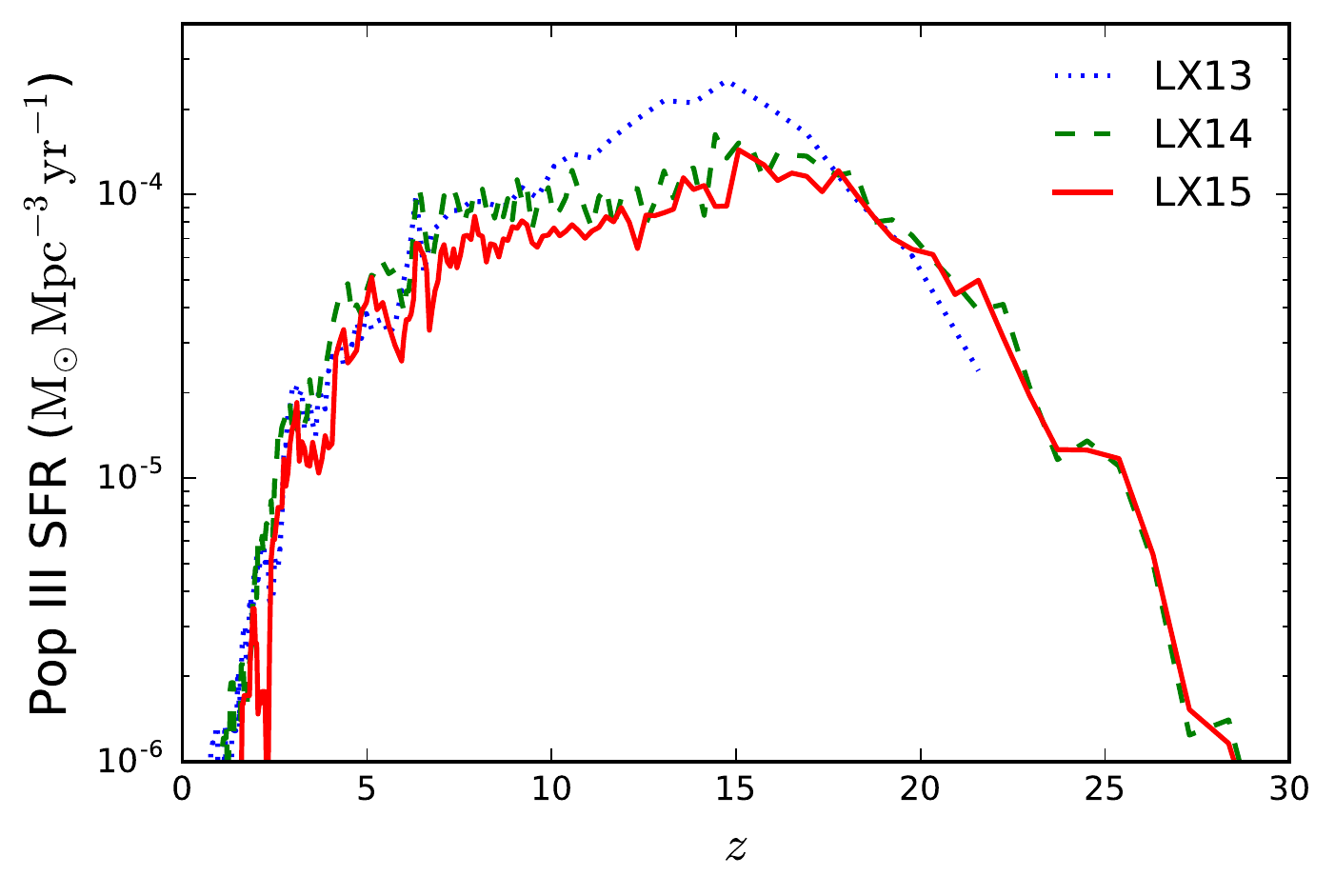}
 \caption{\label{fig:res} Pop~III SFRs for one of the \textit{Caterpillar} haloes at three different resolutions, where LX13 is the lowest and LX15 is the highest resolution. The LX14 resolution run, i.e.\ the resolution used in the main study, produces very similar SFRs to those in the higher resolution run. The LX13 resolution run shows deviations at $z\ga10$. The SFRs have been averaged over $\Delta z=0.2$ for clarity.}
\end{figure}

We test whether the mass resolution of the \textit{Caterpillar} merger trees is sufficient for our model of Pop~III formation. We model Pop~III formation in the same \textit{Caterpillar} halo at three different resolutions: the one used in the main model (LX14, particle mass $2.99\times10^4\,\Ms$), lower (LX13, particle mass $2.38\times10^5\,\Ms$) and higher (LX15, particle mass $3.73\times10^3\,\Ms$). In Figure \ref{fig:res} we present Pop~III SFRs for these three models. The SFRs for the two higher resolutions are very similar, which was to be expected as \cite{Griffen17} have shown the number of minihaloes to be well converged. The LX13 run produces notably different SFRs at $z\ga10$. We conclude that mass resolution used for this project is necessary and sufficient to obtain well-resolved star formation histories. As the LX15 run does not reach $z=0$ we cannot compare the final derived properties of the Milky Way and its satellites for the three resolutions.

\section{Ionization model}
\begin{figure}
 \includegraphics[width=0.99\linewidth]{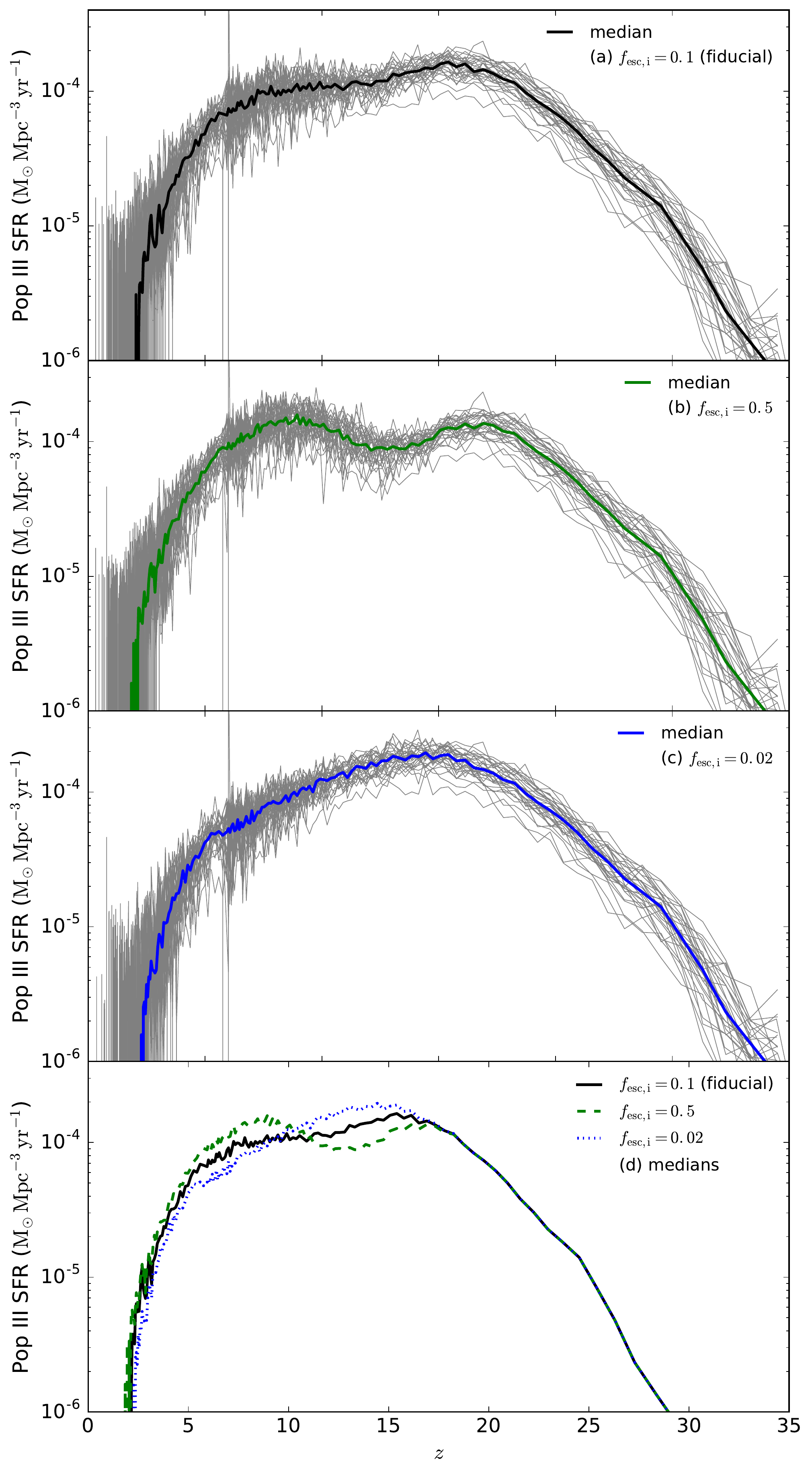}
 \caption{\label{fig:ion} Comoving Pop~III SFR densities for different ionizing photon escape fractions. The shape of the plateau between $5\la z \la 17$ is sensitive to $f_\mathrm{esc,i}$, but the total SFRs are not.}
\end{figure}

Additionally, we test the impact of modifying the details of our ionization model. The IMF, ionizing photon escape fraction, and the star formation efficiency all affect the strength of the ionizing radiation feedback. As representative for modifying the strength of the feedback we vary the ionizing photon escape fraction $f_\mathrm{esc,i}$ by a factor of 5. We present the SFRs for these test runs in Figure \ref{fig:ion}. Decreasing $f_\mathrm{esc,i}$ shifts star formation around from after $z\sim 9$ to before $z\sim 9$. Some of the Pop~III star formation occurs in minihaloes instead of atomic cooling haloes. Therefore, the number of Pop~III forming progenitors (Figure \ref{fig:prog}) increases by 20 per cent. Increasing $f_\mathrm{esc,i}$ has the opposite effect and the number of progenitors decreases by about 10 per cent. The slope stays the same in both cases. There is no notable difference in the number of Pop~III survivors (Figure \ref{fig:surv_law}).

\bsp 
\label{lastpage}
\end{document}